\documentclass[manuscript,screen, acmsmall]{acmart}
\usepackage{wrapfig}
\usepackage{subfigure}
\usepackage{pifont}
\usepackage{enumitem}

\AtBeginDocument{%
  \providecommand\BibTeX{{%
    \normalfont B\kern-0.5em{\scshape i\kern-0.25em b}\kern-0.8em\TeX}}}

\setcopyright{acmcopyright}
\copyrightyear{2022}
\acmYear{2022}
\acmConference{CSCW '22}
\acmISBN{Virtual}

\definecolor{agreen}{RGB}{74, 198, 148}
\definecolor{purple}{RGB}{173, 141,174}

\newcommand{\cmark}{\ding{51} }
\newcommand{\xmark}{\ding{55} }

\begin{document}

\title[Online Payments During the COVID-19 Pandemic]{Six Feet Apart: Online Payments During the COVID-19 Pandemic}

\author{Omar Shaikh}
\authornote{Work completed during an internship at Block.}
\email{oshaikh@gatech.edu}
\affiliation{%
  \institution{Georgia Institute of Technology}
  \country{USA}
}

\author{Cassandra Ung}
\email{cassung@squareup.com}
\affiliation{%
  \institution{Block, Inc}
  \country{USA}
}

\author{Diyi Yang}
\email{diyi.yang@cc.gatech.edu}
\affiliation{%
  \institution{Georgia Institute of Technology}
  \country{USA}
}

\author{Felipe Chacon}
\email{felipe@squareup.com}
\affiliation{%
  \institution{Block, Inc}
  \country{USA}
}

\renewcommand{\shortauthors}{Shaikh et al.}

\begin{abstract}
    Since the COVID-19 pandemic, businesses have faced unprecedented challenges when trying to remain open. Because COVID-19 spreads through aerosolized droplets, businesses were forced to distance their services; in some cases, distancing may have involved moving business services online. In this work, we explore digitization strategies used by small businesses that remained open during the pandemic, and survey/interview small businesses owners to understand preliminary challenges associated with moving online. Furthermore, we analyze payments from 400K businesses across Japan, Australia, United States, Great Britain, and Canada. Following initial government interventions, we observe (\textit{at minimum for each country}) a 47\% increase in digitizing businesses compared to pre-pandemic levels, \textit{with about 80\% of surveyed businesses digitizing in under a week}. From both our quantitative models and our surveys/interviews, we find that businesses rapidly digitized at the start of the pandemic in preparation of future uncertainty. We also conduct a case-study of initial digitization in the United States, examining finer relationships between specific government interventions, business sectors, political orientation, and resulting digitization shifts. Finally, we discuss the implications of rapid \& widespread digitization for small businesses in the context of usability challenges and interpersonal interactions, while highlighting potential shifts in pre-existing social norms.
\end{abstract}

\ccsdesc[500]{Applied computing~Digital cash}
\ccsdesc[500]{Applied computing~E-commerce infrastructure}
\ccsdesc[500]{Applied computing~Online shopping}
\ccsdesc[500]{Applied computing~Secure online transactions}
\ccsdesc[500]{Human-centered computing~Empirical studies in HCI}
\keywords{COVID-19, small business, digital money, e-cash, e-wallets}

\maketitle

\section{Introduction}

\begin{quote}
    \emph{``So going back to March [2020], we’re now in this position where our cash reserves are burning like
wildfire. We needed to keep our talent on payroll but also needed to close due to the risk of COVID.
We had to figure something out, fast.''} - P2
\end{quote}

As COVID-19 infections initially spread around the world, from early to mid 2020, small businesses were forced to re-evaluate how to transact in a rapidly changing environment \cite{pivotPayments}. With multiple regions around the world enforcing business-related restrictions \cite{hale2020variation}, sellers were faced with unprecedented challenges when adjusting their business models to continue taking payments. Furthermore, both COVID-19 infection rates and restrictions varied substantially across regions: Australia enforced strong travel bans and mandatory stay-at-home restrictions despite relatively low COVID-19 rates, while some regions in the United States enforced little to no restrictions in spite of soaring COVID-19 rates. Variations in COVID-19 rates, along with corresponding policy, likely determined if businesses could remain open. 

Online stores offer a potential alternative to businesses affected by COVID-19. By offering services and goods online, businesses can avoid the physical limitations associated with COVID-19 transmission. Analogous works on understanding and augmenting physical experiences with digitization (in response to the pandemic) are far-ranging, looking at education \cite{lee2021show, benabdallah2021remote, remoteLearningPandemic}, independent musicians, theatres \cite{theatresCOVID}, and cattle markets \cite{mim2021gospels}. These works highlight both benefits and challenges associated with moving activities online. 
Small businesses, however, are especially unique, since they already face challenges with digitization. For example, older customers may be apprehensive of foreign digital payment methods \cite{vinesCheque} deployed at these businesses. Furthermore, specific service sectors may have difficulty finding appropriate ways to digitize their services. With the COVID-19 pandemic, some businesses may have been forced to digitize \textit{despite} usability concerns. Small businesses must also independently deploy and solve issues with their digital payment infrastructure, unlike significantly larger online shopping services. Understanding the degree to which different business sectors were affected by the pandemic will help quantify the digitization shift, and may provide insight into better serving future concerns (beyond COVID-19) raised by the necessity to move online.

The adoption of digital payment techniques may also vary based on region-dependent factors. Prior work shows that, although government interventions are effective at minimizing the transmission of COVID-19, adherence to these interventions depends on pre-existing social factors \cite{van2020using}. For example, region and political orientation may both affect adherence to social-distancing orders \cite{mobilePhoneCovid, allcott2020polarization}, further influencing the adoption of digitization. Because inter- and intranational responses to COVID-19 vary significantly, understanding what factors correlate with different increases in digitization also sheds light on the effect of government interventions in the context of small business. Finally, many regions were simply unprepared for COVID-19, having little infrastructure in place to handle interventions for slowing transmission \cite{unpreparedCOVID}.

Motivated by these concerns, our work aims to understand the effects of the COVID-19 pandemic on business digitization, via three research questions. 
We start with a comprehensive analysis of the preliminary reasons behind small business' moving online via qualitative studies. Specifically, we examine if policy interventions and/or COVID-19 caused a digitization shift, while documenting early challenges from businesses. 
\begin{itemize}
    \item[] \textbf{RQ1:} How has the pandemic \textbf{caused} small businesses to digitize, and do they face any preliminary usability \textbf{challenges} with digitization?
\end{itemize}
After establishing causes and challenges associated with small business digitization, 
we then quantify and compare digitization internationally via quantitative methods. Prior international cultural differences associated with adopting technological infrastructure \cite{japanMeiwaku} may reflect on COVID-19 digitization \textit{and} its associated challenges. Therefore, to understand digitization at an international scale, we focus on \textit{when} businesses digitized with respect to COVID-19 surges, and quantify differences in digitization across several countries. %
\begin{itemize}
    \item[] \textbf{RQ2:} Have COVID-19 rates \textbf{predicted} digitization shifts through a large duration (11/2019 - 7/2021) of the pandemic? If so, do these changes differ at an \textbf{international level}?
\end{itemize}
To obtain finer grained relationships affecting digitization, our RQ3 builds on RQ2 by taking a closer look at the effect of specific state-level factors (e.g. policy) and business sectors on digitization, during the initial COVID-19 surge (March 2020). Specifically, we will dive deeper into a case-study of the United States, where a majority of our survey sample resides.

\begin{itemize}
    \item[] \textbf{RQ3:} How have different \textbf{intranational government interventions} and \textbf{regional factors} in the United States affected initial digitization for \textbf{distinct small business sectors?}
\end{itemize}
Overall, understanding how businesses adopted digital infrastructure in response to the pandemic may help identify challenges associated with moving online. Furthermore, characterizing and quantifying the scale of digitization during the COVID-19 pandemic will highlight the potential impact of improving technological infrastructure to support small businesses. Broadly, our work highlights the following implications for the design and adoption of online payment systems:

\begin{enumerate}
    \item Almost all studied regions have seen a significant and rapid increase in the \textit{initial} adoption of digital payment techniques following interventions. Businesses cited proactiveness, with respect to changing interventions, as a reason behind early adoption. The magnitude of adoption also correlates with interventions and pre-existing social factors. 
    \item Rapid digitization can alienate both businesses and customers. Hybrid digitization strategies (e.g. appointments, gift cards)---that combine an in-person and online experience---are a potential solution for various business sectors and customer demographics. 
    \item Businesses, especially in service oriented sectors, strongly value interpersonal interactions with customers. Hybrid payment interactions have also allowed businesses to stay in touch with customers, while offering a personalized experience. Despite significant early adoption, businesses are still experimenting with strategies that promote interpersonal interactions.  
\end{enumerate}

\section{Background \& Related Work}
\subsection{Trends in Payment Digitization}

Although monetary interactions originate from the exchange of physical tokens, transactions are becoming increasingly digital. The advent of online banking, mobile wallets, and eCommerce technologies have catalyzed a transition from transactions that relied largely on cash towards increasingly digital exchanges \cite{pritchardBusesLondon, maurer2015would}. These technologies are enabling users to interact with their personal finances in novel ways. By moving to a digitally grounded payment system, users can easily analyze prior payment history, recover/record transactions \cite{parikhPaperFinance}, and send money to other individuals almost instantaneously \cite{vinesPayOrDelay}. Furthermore, the advent of digitized wallets allows users to decouple themselves from large financial institutions (i.e., ``unbanking''), and rely on digital wallets (like Venmo and Cash App) to store their income \cite{pickens2009windowUnbanked}. Digital payment wallets, for example, are increasing in popularity around the world. Mobile payment systems like M-Pesa are already acting as a bank for some individuals \cite{mas2009designing}. Some mobile payment wallets, like Venmo and Cash App, are also issuing physical cards that allow users to directly utilize their balances in a mobile wallet, further incentivizing a transition away from traditional banking systems \cite{venmoCard, cashappCard}. 

Although research in understanding the adoption speed of digital payment systems remains somewhat limited, work on the active utilization of payment technologies---like QR codes \cite{maurer2013longQR}, NFC \cite{ailisto2009NFCexperiences}, the design and development of digital wallets \cite{socialDigitalWalletTang}, and the use of mobile payment systems \cite{olsenWallet}---have highlighted both technical benefits and challenges associated with adopting payment digitization \cite{moneyworkPerry, maurer2015would}. In our work, we extend the analysis of digitization internationally, looking at relationships between COVID-19 rates and businesses moving online through the entire duration of the pandemic \textbf{(RQ2)}. Furthermore, we aim to measure the rise of digital payments between businesses and consumers caused by COVID-19, focusing on specific periods of time that caused accelerated adoption of digital payment technologies in the United States \textbf{(RQ3)}.

\subsection{Digital Payments and Usability}

A potential drawback of switching from traditional cash payments to digital mediums are challenges associated with usability. \citet{financeWomenPakistan} and \citet{financialInclusionPakistan} highlight how digitization has potential to improve economic empowerment for women in patriarchal communities; however, they also outline current usability problems that hinder the adoption of mobile money in these communities. Finally, digitization itself may also alienate individuals who are used to analog modes of payment \cite{indiaLoanRepaymentAccessibilityoneill2017digital, userChallengesPayments, applesMobilePay}. \citet{vinesCheque}, for example, explores how elderly users find physical cheques a valuable mode of payment, and how the deprecation of cheques in the UK may initially affect general usability. On the other hand, payment providers are quick to emphasize the speed and ease of digital payments: digitized transactions simply require a swipe, tap, or click and are more secure \cite{ondrus2009nfcFast, balan2009mferioFast, lehdonvirta2009virtualFast, moneySecurity}. 
Prior work on understanding the usability limitations of physical payments highlights issues related to accounting (e.g., counting the right amount change) \cite{accessibiltyIndiaMetro}, transportation, and safety. Furthermore, the prevalence of smartphones has enabled low-income communities to take advantage of mobile wallets. Banks---with relatively low penetration, safety concerns, and inconvenience---pose significant usability roadblocks (compared to mobile phones) in these communities \cite{pickens2009banking, penicaud2019state}. Usability benefits also extend to online delivery services (which use digital payments), potentially providing healthier food choices to communities in ``food deserts''---regions where healthy food choices are scarce \cite{groceryDillahunt2019online}. 

These aforementioned works are contextualized in a pre-COVID-19 world. The pandemic, however, has narrowed options for usable payments; because the virus can spread through aerosolization \cite{anderson2020consideration} and potentially via physical currency \cite{moneyDisease}, sellers and buyers may find themselves in situations where in-person payments are not safe, rendering prior payment methods unusable. In some regions, restrictions on gathering size and dine-in restaurants spurred the adoption of digital payment options \cite{pivotPayments}. As a result, payment processing services like Mastercard have noticed a significant increase in digital and contactless payments driven by COVID-19 \cite{mastercardCovid}. To continue making payments usable, sellers may have had to adopt alternative payment acceptance methods; to this end, our work briefly touches on cases where COVID-19 forced some sellers to switch \textit{despite} usability benefits lost from traditional cash transaction methods \textbf{(RQ1, RQ2)}. 

\subsection{International Variations Related to COVID-19}
\label{int_variation_background}
In this work, we study all countries supported by a financial services company that provides optional digitization offerings to small businesses; specifically, we analyze digitization in the US, GB, CA, JP, and AU. However, these countries have key variations that may cause differences in adopting digital payment techniques. In all of these countries, restrictions on gathering, stay at home mandates, business closures, and restrictions on international travel were all implemented in the early stages of the pandemic \cite{hale2020variation}. While the US, GB, CA, and AU enforced these safety measures through fines, JP just had these measures as recommendations \cite{mitigationAUandJP}. Although COVID-19 government interventions generally appeared early, international stringencies have varied significantly through the current duration of the pandemic \cite{hale2020variation}.

From a fiscal perspective, each of the countries we studied had some sort of small business loan available, along with strategies for assistance to those who are unemployed. AU, JP, GB, and CA also had strong employment protection policies in place. For example, AU had the JobKeeper and JobSeeker initiatives that encouraged businesses to retain employees by paying \$1,500 bi-weekly per employee to businesses~\cite{AUPlan}. Canada had the Canada Emergency Wage Subsidy program which covered up to 75\% of an employees’ salary for up to 3 months \cite{canvUSA}. Similarly, GB offered the Coronavirus Job Retention Scheme \cite{pope2020coronavirus}, which initially allowed employers to claim 80\% of an employees salary in grants, or \pounds 2500 per month. Finally, in late April, Japan provided approximately two-thirds of an employees wage if they were placed on furlough. In contrast, the US has the Paycheck Protection Program (PPP), which was targeted towards assisting businesses as a whole, instead of individual employees (i.e. no direct wage subsidy programs) \cite{canvUSA}. 

Finally, international cash circulation rates also varied substantially prior to COVID-19. \citet{thomas2013measuring} estimates that a majority of consumer transactions in Canada (57\%) and United Kingdom (52\%) were completed via ``non-cash'' methods. The United States sees similarly high levels (45\%) of non-cash methods, while Australia and Japan see relatively lower rates (35\% and 14\% respectively). In our work, we suspect that these policy and regional variations may indirectly result in international differences in digitization. Through \textbf{RQ2}, we quantify shifts in digitization and compare these shifts across countries, testing if significant digitization variations exist internationally. 

\subsection{Social Implications of COVID-19}
\label{social_implications}
Prior work raises concerns about current and long-term social implications related to COVID-19, discussing topics like socioeconomic instability, elevated stress levels, disruption in K-12 education, widespread uncertainty regarding the future of COVID-19, and long-term separation from family and loved ones \cite{cao2020psychological, wang2020psychological, psychCovid, nicola2020socio}. A result of the COVID-19 pandemic is increased psychological distress: \citet{PsychologicalDistress} compared levels and symptoms of psychological distress and loneliness in US adults between April 2020 and 2018, finding increases in 2020 related to the spread of the COVID-19 virus. Complementing this work, \citet{saha2020covid} analyzed increased mental health concerns caused by COVID-19, gauging the psychosocial effects of COVID-19 on Twitter compared to 2019. Social implications also apply to computer-mediated interactions: the virtualization of communication raises issues like ``Zoom Fatigue,'' where reported exhaustion has increased due to frequent teleconference interactions \cite{Bailenson2021Nonverbal}. 

Like prior psychosocial work, we also analyze the impact of COVID-19 on behavioral changes. However, we focus on changes caused by the pandemic in the context of digital payment technologies, looking at the immediate effects of initial government interventions in the United States \textbf{(RQ3)}. Because the impact of COVID-19 is region-dependent, we also aim to study digitization internationally, where restrictions have varied throughout the duration of the pandemic; and where COVID cases have increased and decreased at different rates \textbf{(RQ2)}.

\subsection{Strategies for Moving Businesses Online}
\label{online_strategies}
Businesses can utilize various strategies to move their business online. In our work, we contextualize digitization through a pre-defined set of strategies. Although we aim to make these techniques generalizable, we note that they are inspired by services provided by the financial service company studied in this work. These strategies allow sellers to transition much or all of their payment flow to be contactless, minimizing risk of virus transmission. We analyze the adoption of these techniques as a whole; in future work, we hope to understand \textit{specific} scenarios where each of these distinct strategies are best utilized in the context of the COVID-19 pandemic. In this work, however, we measure digitization through the use of only one of the following strategies; specifically, a seller is considered to be ``digitized'' if they adopt at least one of the following in response to the pandemic or a government intervention. 

\subsubsection{Digital Delivery Services} One strategy for offering digital services involves the use of delivering services and food to customers, through a mobile app. Example services that cater to restaurants, like Uber Eats, DoorDash, and GrubHub, have garnered significant attention since the start of the pandemic \cite{mktwtch}. Complementary services that cater to retail, like Instacart, have also gained traction \cite{wsjGrocery}. Sellers can adopt these techniques to offer contact-free delivery. For example, employees of delivery services can provide customers with the option to leave a delivered good at their doorstep; using delivery services allows customers to both minimize in-person interactions and reduce points of contact.

\subsubsection{Online Catalogs and Menus} Another strategy associated with the growth of online commerce is the use of online menus and catalogs. These menus allow users to browse and order products offered at a business in a digital medium. Unlike delivery services, online catalogs and menus can be utilized when physically visiting a store. In some instances, users can also order through the digital menu itself. Sellers can display QR codes that link to menus or catalogs, reducing contact for both employees and customers by eliminating the need to touch a physical menu. 

\subsubsection{Digitally Assisted Curbside Pickup} In contrast to delivery services, some businesses offer curbside pickup, allowing users to pick up goods after ordering them online. Users can place an order online, then visit the business in person, where staff will bring the order to the curbside. Curbside pickup offers sellers a flexible method for continuing to take payments near their physical store, while decreasing opportunities for contact by limiting in-store interactions.   

\subsubsection{Invoicing Digitized Payments} 
Another feature that comes with the above techniques for digitizing payments is the ability to easily track and monitor payment history for sellers and customers \cite{kaye2014money}. All the aforementioned techniques come integrated with the ability to send digital invoices to customers' personal devices, allowing sellers to retrieve, track, and analyze sent invoices. This enables both sellers and customers to record-keep payments in a contactless fashion.

\begin{figure*}
  \centering
  \subfigure[Business Sector]{\includegraphics[scale=0.41]{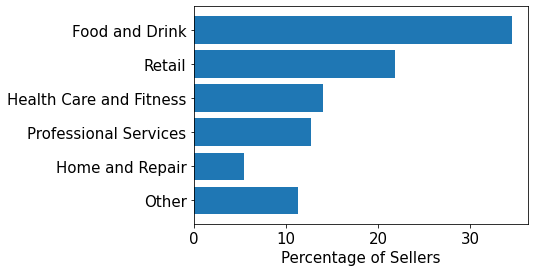}}\hfill
  \subfigure[Country Code]{\includegraphics[scale=0.41]{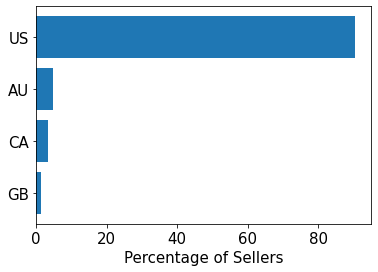}}
  \caption{Distribution of N=993 surveyed businesses across Sector and Country}
  \label{countryFigureSurvey}
  \label{categoryFigureSurvey}
\end{figure*}

\begin{figure*}
  \centering
  \subfigure[Digitization responses caused by the pandemic. The orange bar represents the subset of sellers who digitized specifically because of COVID related restrictions.]{\includegraphics[scale=0.33]{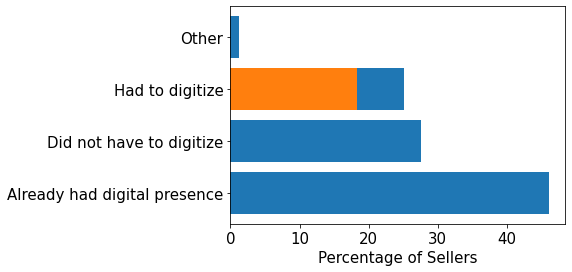}}\hfill
  \subfigure[Speed of digitization for sellers who ``Had to digitize'' because of the pandemic.]{\includegraphics[scale=0.33]{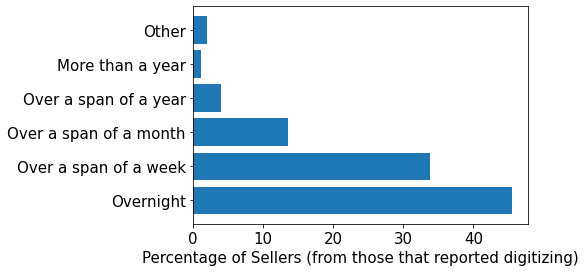}}
  \caption{Distribution of N=993 surveyed businesses on Digital Status and Speed}
  \label{surveySpeedDistr}
\end{figure*}

\section{Dataset Collection Overview}
\subsection{Initial Dataset Source}
\label{data_overview}
To collect our data, we partnered with Block Inc., a financial services company. We specifically use data from Square (a Block Inc. product). Square provides businesses with methods to digitize payments (outlined in Section \ref{online_strategies}). Our research focuses on small businesses using these digitization offerings, with small businesses being defined based on the U.S Small Business Administration (SBA) \cite{sbaTable} as having less than 500 employees and generating less than 7.5 million USD in income \cite{dilger2012small, sbaTable}. When using our data, we followed privacy guidelines established by the company's internal policies, accessing personally identifiable information only when necessary (to contact businesses for surveys and interviews).

From the financial services company, we first identified target businesses that remained active before \textit{and} throughout a year of the pandemic. We also limited our target businesses to countries that were supported by the financial services company during this duration: United States (US), Australia (AU), Great Britain (GB), Japan (JP), and Canada (CA). To identify active businesses, we selected sellers that took more than 1.5 payments per week on average between March 2019 and March 2021, excluding federal holidays. Data on payment frequency was also made available through the financial services company. 

The March 2019 and March 2021 activity time-frame allowed us to select businesses that remained active one year before/after the initial COVID-19 case surge in March 2020 \footnote{Although we select for active \textit{businesses} during this time-range, we still collect \textit{all} payments, even if they were before/after the date ranges used to test for activity}. Our initial sample---after filtering for activity---consisted of about $400,000$ businesses. From each of these businesses, we collected data on the business type, number of daily transactions, and the transaction medium (i.e. if the transaction was completed digitally through the items listed in Section \ref{online_strategies}, completed physically via a card, or through cash). All analysis in this work is derived from these initial businesses---for the next subsections, we cover how we construct subsets tailored towards each RQ.

\begin{table}[]
\renewcommand{\arraystretch}{1.1}
\setlength{\tabcolsep}{2pt}
\small
\centering
\begin{tabular}{@{}ll@{}}
\toprule
Business Category & Definition and Examples \\ %
\midrule
\begin{tabular}[t]{@{}l@{}}Food and Drink \end{tabular}& \begin{tabular}[t]{@{}l@{}} Establishments that primarily serve food or drink. \\ 
\emph{restaurants, grocery stores, bakeries, bars, caterers, food trucks, outdoor markets, etc.}\\  \end{tabular} \\ %

\begin{tabular}[t]{@{}l@{}}Retail \end{tabular}& \begin{tabular}[t]{@{}l@{}} Retail stores that provide goods to customers (primarily non-food related goods) \\ 
\emph{clothing, jewelry, sporting, furniture, flowers, electronics, office supplies, etc.}
\end{tabular}\\ %

\begin{tabular}[t]{@{}l@{}}Health Care and Fitness \end{tabular}& \begin{tabular}[t]{@{}l@{}} Businesses that provide services or goods related to health and general wellness \\ 
\emph{medical services (dentistry, eye care), hospitals, gyms, pharmacies, etc.}
\end{tabular}\\ %

\begin{tabular}[t]{@{}l@{}}Professional Services \end{tabular}& \begin{tabular}[t]{@{}l@{}} Businesses that provide misc. (excluding repair) professional services \\ 
\emph{art and photography, (legal) consulting, accounting, web dev, car washing, etc.}
\end{tabular}\\ %

\begin{tabular}[t]{@{}l@{}}Home and Repair \end{tabular}& \begin{tabular}[t]{@{}l@{}} Businesses that provide goods and services related to maintenance and repair \\ 
\emph{automotive services, heating and plumbing, landscaping, computer repair, etc.}
\end{tabular}\\ %

\begin{tabular}[t]{@{}l@{}}Other \end{tabular}& \begin{tabular}[t]{@{}l@{}} Businesses that do not fall into the aforementioned categories \\ \emph{casual use, transportation, leisure and entertainment, charities, etc.} \end{tabular}\\ %
\bottomrule
\end{tabular}
\caption{Categorization of small business sectors, along with examples that fall into each categorization}
\label{sector_types}
\end{table}

\subsection{Data for RQ1: Surveys and Interviews for Digitization Challenges \& Causes}
\subsubsection{Surveys}
\label{survey_data_section}
Using an email campaign, we reached out to a random sample of 20,000 sellers from the initial $\approx400,000$ (described in Section \ref{data_overview}) that remained active before/throughout the pandemic; email addresses were collected through the financial services company's database.

Consisting of several multiple choice questions, our survey focused on understanding how payments were accepted during the pandemic. We asked about specific methods (from Section \ref{online_strategies}) sellers used in their own business, reasons for switching to digital payment methods, and the speed of the transition itself. We also included optional free response questions asking for further feedback or observations about the speed and challenges associated with business digitization. Survey responses were anonymous at the individual level: respondents were not asked to input information about their name, gender, age, etc. We did, however, collect information about business sector, along with contact emails, directly through the financial services company's internal database. Specific questions, along with details related to the survey platform, can be found in Appendix \ref{fullquestionlist}. Our 993 respondents represented over 848 cities across 74 states/provinces and 4 countries (US, AU, GB, CA). Respondents were distributed across several sectors as defined in  Table \ref{sector_types} and highlighted in Figure \ref{categoryFigureSurvey}. We did not compensate individuals for responding to our survey.

\begin{table}[]
\small
\begin{tabular}{lrrrrr}
\toprule
                                               & Sector & Interviewee Role & Age & Gender & Country \\
\midrule
P1                          &       Retail (Clothing)  &     Marketing Director       &     39  &         F  &  US  \\
P2                          &       Food \& Drink (Dessert)  &      Head of Brand Strategy     &     29  &        M  &  AU  \\
P3                          &       Food \& Drink (Bar)   & Manager \& Co-founder &     38  &         M  &  JP  \\
P4                          &       Prof. Service (Makerspace)    &  Operations Manager   &     38  &         M  &  JP  \\
P5                          &       Food \& Drink (Bar)   &  Co-founder     &     44  &         M  &  CA  \\
\bottomrule
\end{tabular}
    \caption{Participant information for our interviews. Business names are anonymized for peer review.}
    \label{user_stories_table}
\end{table}

\subsubsection{Interviews}
Alongside our survey, we also conducted semi-structured interviews with sellers who were in contact with the financial services company we study. We found these businesses either through their prior contact, or through a hot-line set up for businesses to contact the financial services company. We used a mixture of purposeful and convenience sampling to identify businesses whose characteristics were of interest to our work \cite{robinson2014sampling}. In all, we recruited 5 individuals who ran (or helped run) a small businesses that reported digitizing during the pandemic. These businesses were active for at least a year before the pandemic, and remained open (or reopened) after initial government interventions. To gain a global perspective, we interviewed business from all countries that supported digital payments (AU, US, CA, JP), except for Great Britain. An overview on each business, along with their corresponding spokesperson, can be found in Table \ref{user_stories_table}. 

Each interview was about 30 to 50 minutes, and participants were compensated 50 USD for their time. Interviews were conducted either over the phone (audio-only), or through video conferencing, depending on the preference of the spokesperson. We developed questions (in Appendix \ref{fullquestionlist}) for our semi-structured interviews from our survey findings. During our interviews, we found that our participants often answered these questions before we asked them, so we only posited unanswered questions. Furthermore, if interviewees highlighted important information not included in our questions, we asked for clarifications on those observations. 

\paragraph{Limitations} Our initial survey dataset is subject to limitations: surveyed digitized sellers are potentially more likely to respond to an email, resulting in response bias. Furthermore, our survey was conducted in English, targeted towards English-speaking countries (US, AU, GB, CA), since our clustering methods (Section \ref{clustering}) support only English. Finally, because of our interview sampling technique (convenience sampling), our interviewees are not representative of all small businesses. 

\subsection{Data for RQ2: International COVID-19 Levels and Digitization Rates}
\label{rq2data}
For our each of our countries (Australia, Japan, Canada, Great Britain, and the United States) present in our active businesses, we collected two different data subsets to answer when and how digitization levels changed with respect to COVID-19 rates \textit{across} countries.

\subsubsection{Does COVID-19 Predict Digitization?} \label{rq2_shift_data} First, we aim to quantitatively understand if COVID-19 rates predicted digitization through the duration of the pandemic. We collect daily business digitization rates between 11/2019 and 7/2021, for all selected businesses active during 1 year of the pandemic (discussed in Section \ref{data_overview}): we select this entire time span to cover all past COVID-19 surges, as seen in Figure \ref{covid_rate_corr}. Alongside business digitization share, we collect new 7-day lagging COVID-19 cases. Weekly national COVID-19 case data is sourced from \citet{owidcoronavirus}. 

\subsubsection{Quantifying Initial Changes in Digitization} Next, we aim to understand how large of an initial digitization shift each country experienced, with respect to a control duration. We collect business digitization rates from a smaller timespan, using 11/2018–6/2019 as a control, and 11/2019–6/2020 as a treatment --- 8 non-overlapping months of data for each group. We use these dates since initial interventions and COVID surges occurred at the midpoint of our selected time-frame (in March 2020). In Section \ref{its_strategy_general}, we describe our experimental setup that utilizes this data.

\begin{wrapfigure}{r}{0.4\textwidth}
  \begin{center}
    \includegraphics[width=\linewidth]{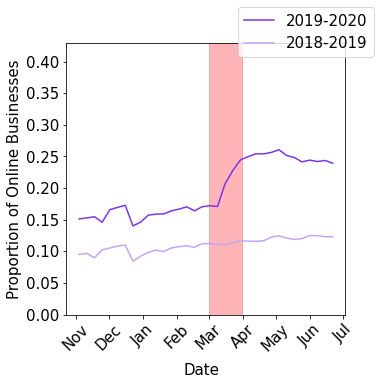}
  \end{center}
  \caption{Data for RQ3: Overall share of businesses that have accepted an online payment at a weekly level in the United States, compared between 2019 and 2020.}
  \label{nationalDigitization}
\end{wrapfigure}

\paragraph{Limitations} For RQ2, we only utilize data at an international level, and do not collect business categories or segment payments by state/province due to limited observations and dataset noise. Furthermore, a potential limitation of our international digitization datasets are a lack of coverage across digitization techniques: in Japan specifically, the financial services company we study supported only digital invoices early in the pandemic (March 2020), and added additional features later. Although this work aggregates all digitization techniques, a potential avenue for future work (discussed in Section \ref{discussionSectionRef}) involves understanding the use of different digitization strategies.

\subsection{Data for RQ3: Digitization Rates and Intranational Factors for The United States}
For RQ3, we use a subset of digitization data similar to \textit{Quantifying Initial Changes in Digitization} from RQ2 (\ref{rq2_shift_data}), but focus entirely on the United States instead of internationally. \textit{Unlike RQ2,} we stratify our data at the state and business category level, since we aim to identify relationships between business sectors and interventions in each state. We limit our analysis to the United States (intranational/statewide level) instead of a international/countrywide level for two reasons. \footnote{We also revisit these reasons/limitations in the discussion (Section \ref{discussion_limitations}).} 

\begin{itemize}
    \item A majority of our sellers ($> 80\%$) from both our survey and our initial set of $400,000$ businesses reside in the United States. Segmenting datasets from other countries at an intrantional level would yield sample-size problems; further segmentation into business sectors would also be prohibitive.
    \item Comprehensive collections of intranational intervention data were limited to a small subset of countries, which included the United States. Intersecting this subset with the list of countries supporting digitization techniques (US, AU, GB, CA, JP) left us with only the United States.
\end{itemize}

\paragraph{State-level Factors \& Intervention Data} After quantifying initial digitization shifts, we aim to explain potential regional factors in the U.S. that contribute to digitization. First, we collect intranational government intervention data: to identify the type and enforcement date of interventions, we utilized the OxCRGT dataset \cite{hale2020variation}. OxCRGT consists of statewide and international data on policy interventions across federal and state governments. The dataset includes 18 indicators that highlight the ``stringency'' of an intervention. Among these indicators, OxCRGT provides several values that affect small business digitization, like government limitations on gathering size, and enforced restrictions on movement (i.e., stay-at-home mandates). We describe a subset of relevant values used in our work in Table \ref{factorRegressionDefs}. Alongside these indicators, the dataset highlights specific dates when indicator values changed---for example, increases in gathering size and business closure values indicate the enforcement of policy interventions. 

We also collect COVID-19 intensities and population density data for each state, in order to control for external effects that may also cause increases in digitization. For COVID-19 intensities, we use data curated by The New York Times \cite{covidDomesticData}, normalizing new cases per 500,000 people in a specified state. Finally, statewide population density data is collected from the 2010 US Census \cite{census:2011}. 

\begin{figure}[t!]
  \centering
          \includegraphics[width=\linewidth]{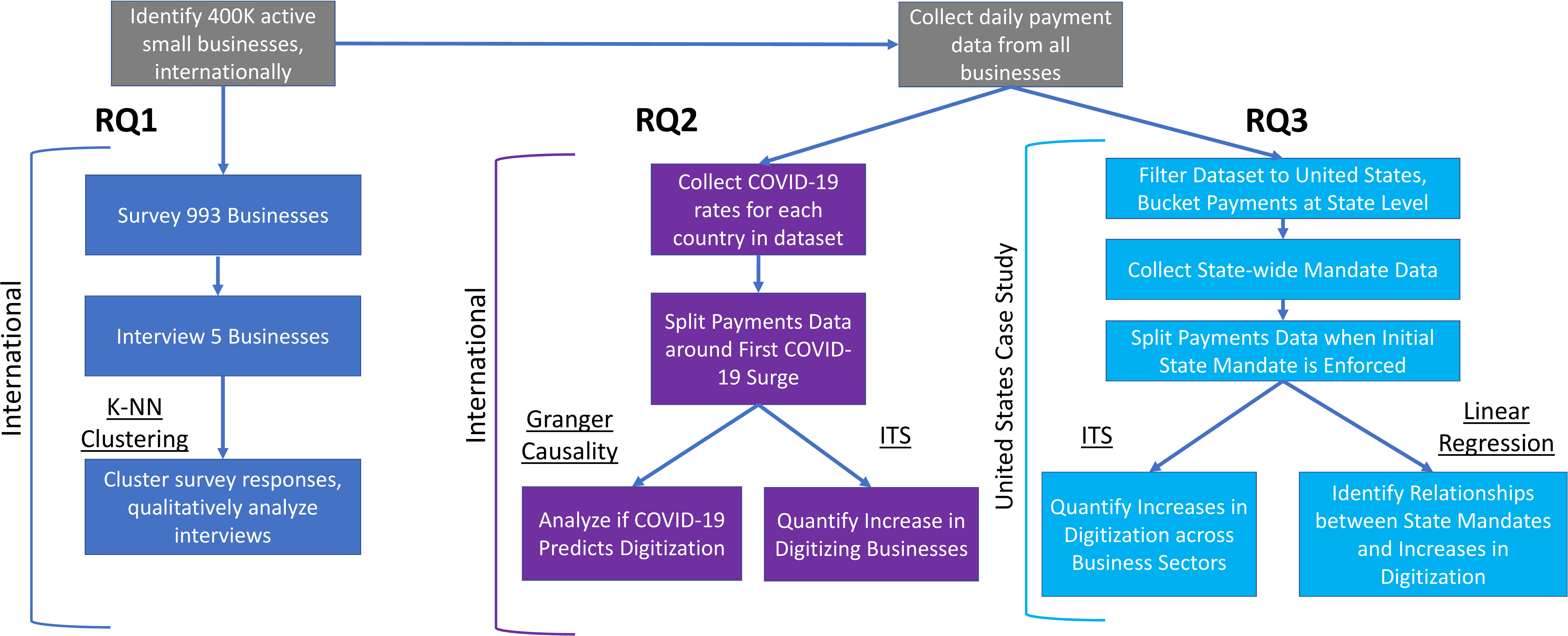}
  \caption{Summary of Data Collection and Methodology. First, we qualitatively identify challenges and causes for digitization (RQ1). Next, we quantify international shifts in digitization and understand how they correlate with COVID-19 (RQ2). Finally, we focus on the United States to better understand the role of regional factors (interventions, political orientation, business sectors, etc.) on digitization (RQ3).}
  \label{methodology_figure}
\end{figure}

\section{Methods}

Figure \ref{methodology_figure} summarizes our methodological approaches; at a high level, we surface challenges and causes related to digitization through surveys and interviews (RQ1), quantify and compare digitization at an international scale (RQ2), and understand the effects on digitization across policy and business sector for small businesses in the US (RQ3). In all, our methods aim to paint a broad quantitative and qualitative picture of the digitization shift catalyzed by COVID-19. %

\subsection{RQ1: Identifying Challenges and Causes Related to Digitization}
\label{clustering}
\paragraph{Clustering Open-ended responses} To identify broader themes in our survey, we cluster open-ended responses. 38\% (380) of our 993 respondents provided open-ended responses to questions in our survey. To identify main topics associated with these open-ended responses, we first automatically cluster responses using natural language processing techniques. For our clustering techniques, we converted each open-ended response into a semantic representation vector, using SentenceBERT \cite{reimers-2019-sentence-bert}. Next, we clustered these vectors with KMeans clustering, using SSE (sum of squared errors) to evaluate various values of $k$ (Figure \ref{clustering_sse}, in Appendix). Using $k=15$, we identified initial themes and manually recategorized misclassifications, merging clusters into 4 broader groups.

\subsection{RQ2 \& RQ3: Quantifying Digitization with an Interrupted Time Series Model}
\label{specific_its}
For both RQ2 \& RQ3, we aim to quantify the shift in digitization. Though RQ2 focuses on digitization across countries (internationally), and RQ3 focuses on digitization in the United States (intranationally), we use the same quasi-experimental methodology to identify if digitization shifts (during 11/2019–6/2020) were significant compared to a control time-span (1/2018–6/2019). Prior work on understanding effects on social media for COVID-19, along with research on applying interrupted time series models to study social network activity, motivated our approach \cite{saha2020covid, redditBan, sahaStress, chandrasekharan2020quarantined}. Furthermore, utilizing a control dataset at a region level allows us to control for geographic factors and seasonality trends within these regions, adjusting for external effects on payment transactions. By highlighting the effects of digitization on an experimental and control time span, we can make claims about the effect of COVID-19 on the rate of digitization.

\paragraph{Bucketing Payments}
First, we bin payments from sellers in a selected region into daily buckets---where each bucket corresponds to a single set of counts. Businesses are counted as ``digitized'' at a specific point in time, if they have taken at least one payment per day using a strategy discussed in Section \ref{online_strategies}. These strategies are also the same as those covered in our Background section and our surveys (delivery services, curbside pickup, online catalogs and menus). We also split payment digitization rates into two periods: $P_{pre}$, and $P_{post}$, indicating the payments before or after a split point. For RQ2, we select March 1 as a split point, since interventions generally appear in these countries after this date \cite{hale2020variation}. For RQ3, we utilize our collected intervention data to identify a split point at the first COVID-19 related intervention in a state.

\paragraph{General Interrupted Time Series Model}
\label{its_strategy_general}
We aim to compare digitization interruptions across regions (countries and states) to understand if some countries digitized significantly more than others in response to the pandemic. To quantify the initial digitization shift, we utilize an interrupted time series (ITS) strategy~\cite{bernal2017interrupted} to identify if business digitization increased significantly following an intervention. Using ITS allows us to account for changes due to policy interventions even in the presence of temporal trends. Importantly, ITS regresses on temporal variables from the dataset as part of the inference process, instead of assuming independence between pre- and post-intervention periods. ITS uses a binary variable $\delta(t>\tau)$ to indicate if a data point $y_{t}$ occurs after a post intervention period (i.e., if $y_{t}$ falls into $P_{pre}$ or $P_{post}$). In our ITS models, $y_{t}$ corresponds to the count of digitizing sellers at day $t$. For our underlying probability distribution function $f$, we use the Poisson distribution; however, we include an offset term $log(N)$ to account for the total number of active businesses $N$ in a given day $t$. We include $a_t$ to regress on time, adjusting for potential seasonality trends. Finally, $z$ indicates if the data belongs to the control or treatment group.

\paragraph{RQ2 International ITS} For RQ2's ITS model, we include interaction terms to model increases across countries $c$.

\begin{align}
y_{t} &\sim a_{t} + \delta(t>\tau) \cdot z \cdot c \\
log(y_{t}) &= \beta_{0} + \beta_{1} \cdot a_{t} + \beta_{2} \cdot \delta(t>\tau) + \beta_{3} \cdot \delta(t>\tau) \cdot z + \beta_{4} \cdot \delta(t>\tau) \cdot z \cdot c  + log(N)
\end{align}

\paragraph{RQ3 United States ITS} For RQ3's ITS model, we extend the model in RQ2 to include shifts across $b$ \textit{business categories}, and replace $c$ country for $s$ state to model intranational variations.

\begin{align}
y_{t} &\sim a_{t} + \delta(t>\tau) \cdot z \cdot s \cdot b \\
log(y_{t}) \nonumber &= \beta_{0} + \beta_{1} \cdot a_{t} + \beta_{2} \cdot \delta(t>\tau) + \beta_{3} \cdot \delta(t>\tau) \cdot z \: + \\ 
&\quad  \beta_{4} \cdot \delta(t>\tau) \cdot z \cdot s  + \beta_{5} \cdot \delta(t>\tau) \cdot z \cdot s \cdot b + log(N)
\end{align}

\subsection{RQ2: Highlighting General International Digitization Trends}
\paragraph{Granger Causality: Using COVID-19 Rates to Predict Digitization}
\label{granger}
To understand if COVID-19 levels and digitization rates correlate internationally, we use Granger's causality analysis \cite{geweke1984measures}, a technique grounded in econometrics. Granger's causality assumes that if a time series $X$ causes another time series $Y$, then changes in $X$ will occur consistently ahead of changes in $Y$. If lagging time series $X$ behind $Y$ results in $X$ predicting changes in $Y$, Granger's causality test will highlight a significant correlation between $X$ and $Y$.  Granger's causality test also assumes that time series data is stationary (i.e., does not have underlying seasonal trends); we utilize the Dickey-Fulller test \cite{dickey1979distribution} to ensure our tested time series data is stationary. Finally, correlation from Granger's test does not imply causation; instead, it indicates that one time series dataset has predictive information regarding another time series dataset.

In the context of testing RQ2, we look for correlations between weekly changes in COVID rates ($X$) and changes in share of digitized businesses ($Y$). To obtain $X$ and $Y$, we transform our original time series data, which consists of COVID cases / M and digitization share at day $t$. To account for weekend trends, we smooth our digitization data: first, we group our data by week $w$, then use a difference transformation (subtracting week $w-1$ from week $w$). Our time series data then represents the weekly \textit{change} in COVID cases or digitization. An added benefit of grouping and the difference transformation is a potential removal of seasonality trends, which satisfies a requirement of the Granger causality test. Results from the Dickey-Fuller test show that all time series pairs $X$ and $Y$ for our selected regions have $p < .05$, reinforcing that our transformed data is stationary (without seasonality) and appropriate for Granger's causality test.   

\begin{table}[]
\renewcommand{\arraystretch}{1.1}
\setlength{\tabcolsep}{2pt}
\small
\centering
\begin{tabular}{@{}ll@{}}
\toprule
\textbf{Dependent Variable} & Definition \\
\midrule
\begin{tabular}[t]{@{}l@{}}Post-mandate Digitization\end{tabular} & 
\begin{tabular}[t]{@{}l@{}}Proportion of digital businesses in a state after initial intervention. \\  
\emph{[0, 1]; Avg proportion, calculated from initial intervention to 6/2020.}
\end{tabular}\\ %
\midrule
\textbf{Independent Variables} & Definition \\ %
\midrule

\begin{tabular}[t]{@{}l@{}}Workplace Closing\end{tabular} & 
\begin{tabular}[t]{@{}l@{}} Mandated closing of workplaces  \\ 
\emph{[0..3]; 0 = no measures; 3 = closing for all but essential.}
\end{tabular}\\ %

\begin{tabular}[t]{@{}l@{}}Restrictions on Gatherings\end{tabular} & 
\begin{tabular}[t]{@{}l@{}} Mandated restrictions on gathering sizes  \\ 
\emph{[0..4]; 0 = no measures; 4 = restrictions on $<10$ people. }
\end{tabular}\\ %

\begin{tabular}[t]{@{}l@{}}Restrictions on Internal Movement\end{tabular} & 
\begin{tabular}[t]{@{}l@{}} Restricted movement between cities/regions, incl. stay-at-home \\ 
\emph{[0..2]; 0 = no measures; 2 = Mandatory internal mvmt restrictions. }
\end{tabular}\\ %

\begin{tabular}[t]{@{}l@{}}Facial Coverings\end{tabular} & 
\begin{tabular}[t]{@{}l@{}} Policies on wearing face coverings outside of house. \\ 
\emph{[0..2]; 0 = no measures; 4 = required at all times outside of house. }
\end{tabular}\\ %

\begin{tabular}[t]{@{}l@{}}\textit{Republican Vote Share}\end{tabular} & 
\begin{tabular}[t]{@{}l@{}} \% of population voting for Republican nominees in 2020 election \\ 
\emph{[0, 1]}
\end{tabular}\\ %

\begin{tabular}[t]{@{}l@{}}\textit{COVID Intensity}\end{tabular} & 
\begin{tabular}[t]{@{}l@{}}Avg. number of cases 7 days before inital mandate, per 10,000 people\\ 
\emph{[0, n]}
\end{tabular}\\ %

\begin{tabular}[t]{@{}l@{}}\textit{Weighted Population Density (WPD)}\end{tabular} & 
\begin{tabular}[t]{@{}l@{}}Pop. density weighed by proportion of pop. density in census tracts\\ 
\emph{[0, n]; $WPD_{S} = \sum^{}_{t \in T_{S}}p_{t} \cdot d_{t}$; $T$ = Tracts, $S$ = State}
\end{tabular}\\ %

\begin{tabular}[t]{@{}l@{}}\textit{Pre-mandate Digitization}\end{tabular} & 
\begin{tabular}[t]{@{}l@{}}Proportion of digital businesses in a state before initial intervention. \\  
\emph{[0, 1]; Avg proportion, calculated from 11/2019 to initial intervention.}
\end{tabular}\\ %

\bottomrule
\end{tabular}
\caption{Definitions of variables used in our linear regression to understand state-level factors that contribute to digitization (Section \ref{linreg_def}). Each of these variables communicates a different type/level of restriction, or relevant contextual factor. \textit{Italicized control variables} are not included in the original OxCRGT dataset. Our regression is formulated as: $\mathrm{Post Mandate Digitization} = \alpha + \sum_{n \in \mathrm{factors}}\beta_{n}\mathrm{Factors}_{n}$.}
\label{factorRegressionDefs}
\end{table}

\subsection{RQ3: Interventions and Digitization in The United States}
\label{linreg_def}
\paragraph{Linear Regression: Identifying Relationships between State-level Factors and Digitization} 
Government interventions consist of various factors that make up the \textit{stringency} of the response itself. To understand what elements of stringency may have played a role in the initial set of mandates outlined by state-level governments, we utilize subfactors from the OxCRGT dataset as \textit{independent variables} to predict digitization rates in a state. All variables are documented in Table \ref{factorRegressionDefs}.

Specifically, we select values for each regression variable (Table \ref{factorRegressionDefs}) from the initial week that mandates were enforced in a selected state. Our dependent variable is the average digitization rate in a state from the initial mandate to 6/2020. As part of the regression, we also include \textit{control variables} for the political orientation of a state (using Republican vote-share during the 2020 election), the COVID intensity at the time an intervention was enforced, and the weighed population density of a state. Prior work indicates that adherence to mandates may be dependent on the political polarization in the United States \cite{allcott2020polarization, makridis2020real}; by including political orientation, we aim to understand if this effect also exists for digitization. For COVID intensity, we include a predictor for the average number of cases in the last 7 days immediately before the mandate per 10,000 people, sourced from The New York Times \cite{covidDomesticData}. We also include a log-scaled weighed population density predictor as another control. Instead of using the raw population density from the census data, we weigh population density by population concentration in each census tract at a state-level, following methodology from how the US Census analyzes metro-level population concentrations to understand urbanization \cite{wilson2012patterns}. Using weighed densities allows us to account for potential population density skews caused by states with large, sparsely populated rural areas. Finally, we re-scale all bounded variables to between $[0, 1]$, log-scale unbounded variables, and perform linear regression on these variables to predict the average post-mandate for a state.

\section{RQ1 Results: Business Digitization Survey \& Interviews}
In this section, we answer RQ1---identifying challenges and causes associated with digitization---through a survey analyses and interviews with small businesses.

\subsection{Survey Findings}
\paragraph{\textbf{The pandemic catalyzed the transition to digitization for a significant proportion of sellers.}} 25.2\% of sellers from our survey responded that the pandemic forced them to find methods to digitize their payment transactions. Excluding sellers that \textit{already had a digital presence}, businesses that moved online due to the pandemic accounted for 46.6\% of the responses. Several respondents, who did not rely on digitization prior to the pandemic, highlighted the usefulness of adopting digitization techniques:

\begin{quote}
"\emph{Digital invoicing has been a big help! I'm able to send the customer a digital invoice after service is performed then they pickup their system from our secured weather proof pickup box outside our door.}" - Retail    
\end{quote}

One respondent noted that they would've moved to online payments earlier during the pandemic given its success at their business.

\begin{quote}
    \emph{``I would have launched an online order site earlier if I would have predicted it's success and had the time.''} - Food \& Drink
\end{quote}

We also asked users how quickly they had to digitize due to the pandemic, and if a stay-at-home mandate was the cause of their transition. \textit{Excluding businesses that already had a digital presence}, 79.3\% noted that they needed to switch payment transaction methods in less than a week (Figure \ref{surveySpeedDistr}); 22.7\% reported that a stay-at-home mandate \textit{specifically} caused them to digitize payments. Finally, of businesses that adopted new techniques because of the pandemic, 78.4\% indicated that they would continue to offer digital payment services after the pandemic was over.

\paragraph{\textbf{Although some businesses were affected by the pandemic, others were not required or able to change much about their payment flow to remain open.}} One respondent observed that the lack of change was due to the business sector; some sectors, like the medical field or service-oriented businesses, may not require as drastic a shift as other sectors: \emph{``Medical field, nothing really changed other than putting payments in my hand.''} Several other respondents whose businesses were already working in services like tele-medicine or online mental health consulting highlighted a similar trend, while other businesses started offering online counterparts for the in-person offerings. 

\begin{quote}
    \emph{``I own a[n] [online] mental health practice and the pandemic increased the utilization of our services.''} - Health Care \& Fitness
\end{quote}

Because some business sectors are tightly coupled to digitized or non-digitized techniques, changing payment methods may be a significant challenge. For service sectors, adjusting to a digital medium posed usability problems to potential customers.

\paragraph{\textbf{Some businesses already had a digital presence that grew in usage due to the pandemic}} 46.1\% of all respondents already had a digital presence set up before the pandemic, utilizing a spectrum of digitization techniques. One respondent highlighted their transition from physical to digital receipts/invoices. 

\begin{quote}
     \textit{``I use the invoice feature and the online store a lot more now...it has allowed us to be completely contactless.''} - Retail
\end{quote}
Another respondent also highlighted an increased use in digitization; however, they highlight usability problems related to adopting digitized payments for older individuals, an observation reflected in \citet{vinesCheque}. To remain contactless, sellers might be required to teach users how digital or online payments work. 
\begin{quote}
    \textit{``We were already adopters of digital/card payments before the pandemic, so it was not a challenging transition for us. However, many older customers needed to be educated; that was the biggest challenge.''} - Retail
\end{quote}
Although digitization is promoted as an option for making payments more secure, some businesses noted that customers have a general distrust of taking payments online.
\begin{quote}
    \emph{``Still some members of the public are worried of scams and take online payments as a risk. [ANON] appears very secure but we have to adapt when clients don't want to ``risk'' online payments.''} - Retail
\end{quote}
In general, we noticed that some respondents simply switched over to their pre-existing digital infrastructure. A few sellers mentioned that having these options set up allowed for a transition into a post-pandemic payments world. 
\begin{quote}
    \emph{``Having [digitization] in place already enabled us to continue seamlessly into the pandemic. We already offered delivery, shipping and pickup, so, fortunately for us, we were already ahead of that curve.''} - Food \& Drink
\end{quote}

\paragraph{\textbf{Businesses relied on several digitization techniques to remain online, and most plan to continue using them after the pandemic.}} Of the first three digitization strategies \footnote{We did not survey for Digital Invoicing, since using any of the other strategies implies use of invoicing.} documented in Section \ref{online_strategies} (Digital Delivery Services, Online Catalogs and Menus, and Digitally Assisted Curbside Pickup), 54.2\% of businesses relied on more than one strategy. Furthermore, of businesses that started using digitization because of the pandemic, 86.3\% of businesses plan to continue using at least 1 strategy that they adopted beyond the COVID-19 pandemic; that is, businesses that previously didn't use digital payment acceptance techniques will offer them after the pandemic.  

\begin{table}[]
\small
\begin{tabular}{lrrrrr}
\toprule
Theme                                               & P1 & P2 & P3 & P4 & P5 \\
\midrule
Started digitizing during COVID-19 (limited online presence before)  & \xmark & \xmark & \cmark & \cmark & \cmark  \\
Experimented with online product line throughout COVID-19 & \cmark & \cmark & \cmark & \cmark & \cmark \\
Experienced usability issues with older individuals & \cmark & \cmark & \xmark & \cmark & \cmark \\
Highlighted interventions as a direct cause for digitization & \cmark & \cmark & \cmark & \xmark & \cmark\\
Will continue using digital payment techniques after COVID-19 & \cmark & \cmark & \cmark & \cmark & \cmark \\
Digital offerings will become main source of revenue & \xmark & \xmark & \xmark & \cmark & \xmark \\
\bottomrule
\end{tabular}
    \caption{Summary of themes in interviews.}
    \label{theme_table}
\end{table}

\subsection{Interview Findings}

\paragraph{\textbf{Older and Younger Generations.}} All our participants \textit{independently} highlighted a divide between younger and older generations when adopting digital payment technologies because of the pandemic. P1 highlighted issues related to trust, along with general usability concerns raised by older individuals when using eCommerce sites.
However, alongside these concerns, P1 highlighted how digitization also allowed her to offer appointment oriented services (scheduled digitally), catering to clientele who preferred an in-person experience when shopping for clothes.

P2 also highlighted an analogous problem for dining experiences -- delivery services are often challenging for older individuals to navigate. 
In these cases, both P2 and P4 discussed instances where younger relatives would order online on behalf of older individuals as a form of gifting.
Business sectors, like in our survey, also appeared to affect how businesses moved online. P3, for example, talked about how his specific sector (a bar serving craft beer) attracted a younger customer demographic, resulting in reduced usability issues for his online store.

Beyond business sector, P3 also independently raised Japan's younger generation's rapidly changing cultural norms as a factor behind changes in digitization. 
\begin{quote}
\emph{``Before the pandemic, [digitization] already started with credit cards. You could see [our customers] felt cool pulling out that card, which was a new thing. I saw this more and more: like guy on a date. It's like boom, card -- that was the cool move. COVID just accelerated that trend here.''} - P3
\end{quote}
When we asked P3 how these norms shifted over time, P3 highlighted how digital payment techniques were starting to become more convenient over the last decade, and were integrating useful point systems. We revisit these qualities---digitizing social norms through online gifting, point systems, etc. during COVID-19---in our discussion. 

\paragraph{\textbf{Interventions and Proactiveness.}} Digitization occurred immediately at the start of the pandemic, for all our interviewed businesses. Following initial restrictions and surges in COVID-19 cases, moving online was not a gradual process. Interviewed businesses shifted online immediately during the first surge, in preparation for future uncertainty. P2, for example, highlighted how moving online was a robust solution to the 2nd lockdown in Australia. 

\begin{quote}
    \emph{``It’s not enough for a business to operate as usual. You always have keep finding ways to adapt,
grow and challenge the norm. We asked
– what will this mean for us, not only in the current landscape, but in the future “normal”, whatever that
might be.''} - P2
\end{quote}
While discussing online payments, participants also highlighted how online options complemented (and never replaced) their in-store experience; moving online helped businesses extend their offerings and prepare for future uncertainty: \emph{``[Our online store] hasn't become our main line of business. I mean, we're a bar, by any means. But it's definitely enabled us to do things that we'd normally be able to do without it.''} - P3.

\paragraph{\textbf{Experimenting with Going Online.}} For some of our interviewed businesses, the role of online stores and digital payments were and are still in flux. Although businesses digitized early in response to interventions, their offerings shifted in response to (1) what was scalable online and (2) what customers wanted while in lockdown. During P5's interview, P5 brainstormed ideas on digitizing social interactions through a podcast, replicating some of the in-person characteristics of a bar. 

All interviewed businesses discussed how their catalogs were experimentally refined during the duration of lockdowns and stay-at-home orders. They specifically highlighted how, despite experimentation, there's a distinctly different experience between the online store and the in-person experience. Specifically, P1 and P5 mentioned how in-person experiences allow businesses to easily personalize a shopping experience to a given customer, building loyalty. P1 also discussed how scheduling in-person appointments online is a potential middle-ground between both experiences. Alongside scheduling appointments, P1 discussed her interactive online store, where she integrating 3D maps of her physical business in her catalog. Beyond experimenting with online catalogs, P3 tried different digitization techniques, offering delivery at the start of the pandemic. However, logistical issues with delivering food resulted in P3 scheduling pickups instead.

Although 4 of our 5 participants indicated that online stores served to complement in-person interactions, P4 pivoted his Makerspace from offering in-person services to goods on an online store. He mentioned that his online store may become a main source of revenue following the pandemic, even when full in-person operations resume. 

\section{RQ2 Results: International COVID-19 Digitization}

\label{beyond_section}
In this section, we aim to analyze relationships between COVID-19 rates and international differences in digitization (RQ2), highlighting the scale of observations from RQ1. To this end, we ask if COVID-19 predicts digitization from all regions in our dataset between 11/2019 and 7/2021. Then, we quantify the initial digitization shift in these regions, comparing differences internationally. 

\begin{figure}[h]
  \centering
          \includegraphics[width=\linewidth]{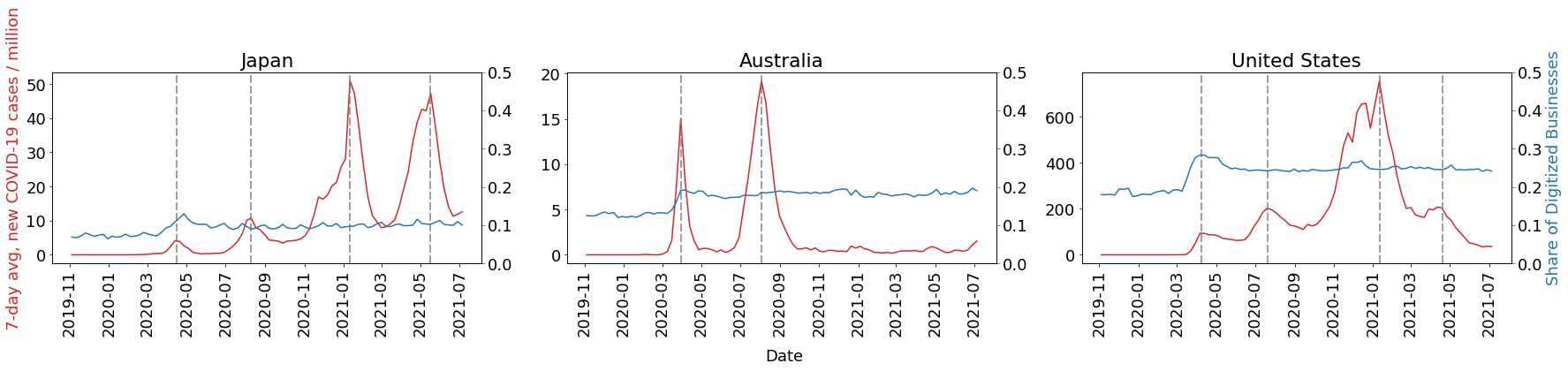}
          \includegraphics[width=\linewidth]{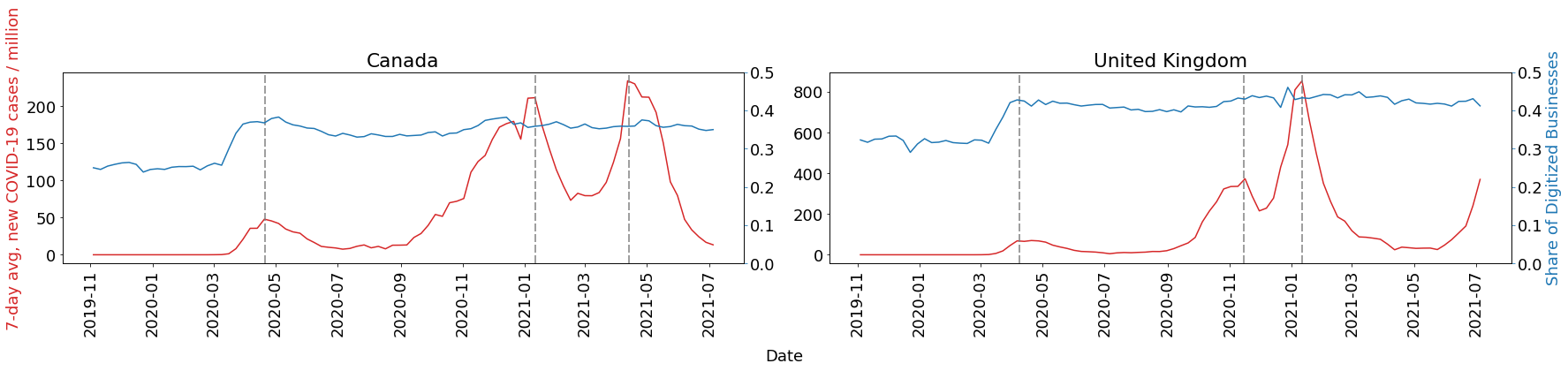}
  \caption{COVID-19 cases / M people and Share of Digitized Businesses from all Supported Countries. Selected Local Maxima for case rates are indicated using a dashed grey line. Qualitatively, changes in COVID-19 rates do not appear to correlate with changes in digitization (aside from 1st surge).}
  \label{covid_rate_corr}
\end{figure}

\newcommand{\ra}[1]{\renewcommand{\arraystretch}{#1}}
\begin{table*}\centering
\ra{1.0}
\begin{tabular}{@{}rrrcrrcrrcrrcrr@{}}\toprule
& \multicolumn{2}{c}{JP} & & \multicolumn{2}{c}{AU} &
 & \multicolumn{2}{c}{US} &  &  \multicolumn{2}{c}{CA} &  & \multicolumn{2}{c}{GB}\\
\cmidrule{2-3} \cmidrule{5-6} \cmidrule{8-9} \cmidrule{11-12} \cmidrule{14-15}
$lag$ & $p$ val. & $F$ stat. && $p$ val. & $F$ stat. && $p$ val. & $F$ stat.  && $p$ val. & $F$ stat.  && $p$ val. & $F$ stat. \\ \midrule
$1$ & 0.24 & 1.39 && 0.11 & 2.58 && 0.82 & 0.05 && 0.33 & 0.96 && 0.63 & 0.23 \\
$2$ & 0.36 & 1.03 && 0.14 & 2.05 && 0.96 & 0.05 && 0.38 & 0.99 && 0.84 & 0.18 \\
$3$ & 0.58 & 0.65 && 0.15 & 1.81 && 0.83 & 0.30 && 0.63 & 0.58 && 0.68 & 0.51 \\
$4$ & 0.51 & 0.83 && 0.14 & 1.72 && 0.94 & 0.20 && 0.79 & 0.43 && 0.81 & 0.40 \\

\bottomrule
\end{tabular}
\caption{Results for Granger causality test on changes in new COVID cases / M ($X$) causing changes in businesses digitizing ($Y$), for varying lags (1 week to 4 weeks) applied to $X$. No setting for lag yields statistically significant results at reasonable $p$ values. Therefore, we fail to reject our null hypothesis (lagged COVID-19 cases do not explain changes in digitization).}
\label{granger_table}

\end{table*}

\label{its_initial}

\subsection{COVID-19 Rates and Digitization Through the Duration of the Pandemic} Since the onset of the COVID-19 pandemic, rates of COVID have varied both over time and across regions. Qualitatively, this observation can be seen in Figure \ref{covid_rate_corr}, where local maximas in COVID-19 rates generally do not correspond to increases in digitization. Granger's causality test quantitatively confirms these observations: results on time series pairs representing changes in COVID rates and digitized businesses are shown in Table \ref{granger_table}. We fail to reject our null hypothesis (lagged COVID-19 cases do not explain changes in digitization); for any setting of lag between 1 to 4 weeks, COVID-19 rates are \textit{not} predictive of digitization share through the current duration of the pandemic. Our Granger Causality test indicates that COVID-19 fluctuations occurring throughout the pandemic \textit{do not} affect digitization. From these results, \textit{we suspect that sellers digitized early in the pandemic.} As highlighted in our surveys \& interviews, when COVID cases surged again, sellers already had digital infrastructure to handle accepting payments. In Section \ref{US_analysis}, we revisit relationships between COVID-19 rates and digitization, analyzing payments in the U.S.

\subsection{International Variations on Digitization}

\label{int_comparision}
We hypothesized that cultural and policy variation would have an effect on digitization internationally: different countries are more/less digitally averse than others. As mentioned in our background (Section \ref{int_variation_background}), a noteworthy example is Japan: historically, Japan's dependence on cash has remained significantly higher than most countries' \cite{cashReport}, so we expected a lower digitization rate in this region compared to others. Qualitatively, we find this trend reflected in digitization \textit{despite} changes in COVID-19 rates throughout the pandemic (Figure \ref{nationalDigitization}). Throughout the pandemic, digitization in Japan appears lower than all other countries, even though COVID rates are generally higher in Japan than in Australia (even though AU saw a higher shift in digitization). Even though Japan's digitization rates are lower, we still see nearly a $50\%$ increase in digital businesses from our ITS model, compared to our 2018-2019 control.

\begin{wrapfigure}{r}{0.4\textwidth}
  \begin{center}
    \includegraphics[width=\linewidth]{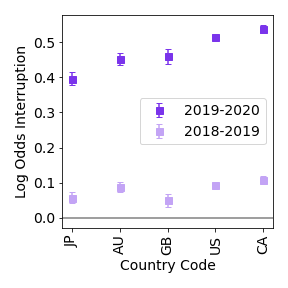}
  \end{center}
  \caption{Increase in Log Odds ($\gamma$ from all state-level ITS Models) after March 2020 for US, AU, and JP. All countries are significantly different from one another (JP < AU = GB < US < CA, $p < .025$, two-tailed $\gamma$ from ITS).}
  \label{int_its_graph}
\end{wrapfigure}

In Figure \ref{int_its_graph}, we highlight the summed interaction parameters (from our ITS model) for the interruption in JP, AU, US, GB, and CA, along with corresponding confidence intervals. Converting the $\gamma$ log-coefficients to relative \% changes (using an $e^\gamma$ transform) results in a 48.64\%, 57.2\%, 58.3\%, 67.4\%, 71.12\% increase in digitization for JP, AU, GB, US and CA respectively. Each of these relative changes have statistically significant differences except for AU and GB, with Japan having the smallest interruption and the CA having the largest  (compared to pre-intervention time spans.)

\section{RQ3 Results: United States Case Study, Initial Digitization}
\label{US_analysis}
In Section \ref{beyond_section}, we looked at international digitization, finding that businesses generally digitized early in the pandemic (instead of alongside COVID-19 surges), and at significantly different rates. In this section, we dive deeper into early digitization, quantifying increases in the United States following initial COVID-19 interventions. We use an interrupted time series technique \cite{bernal2017interrupted} (Section \ref{specific_its}) alongside a linear regression (Section \ref{linreg_def}) to identify the effect of noteworthy regional factors (i.e., stay-at-home mandates, restrictions on gatherings, political polarization; subset described in Table \ref{factorRegressionDefs}) on the adoption of online payments. 
Finally, we highlight differences in digitization across business sectors.

\begin{figure}[h]
  \centering
          \includegraphics[width=\linewidth]{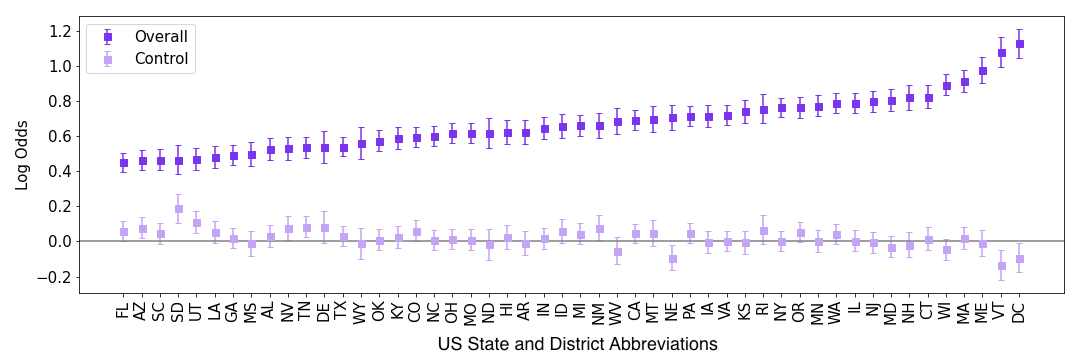}
  \caption{Increase in Log Odds ($\gamma$ from all state-level ITS Models) after March 2020 Business and Gathering Mandates for all business sectors, compared to March 2018–2019 control group. All states had statistically significant differences ($p < .025$, two-tailed $\gamma$ from ITS) for overall digitization compared to the control group. 
  }
  \label{stateVariation}
\end{figure}

\subsection{Increases in Digitization Following Mandates}
\label{gamma_increases_section}
Following government intervention, a significant number of sellers actively began using digitization to continue taking payments. Qualitatively, this trend is visible on a national level: Figure \ref{nationalDigitization} highlights an increase in share of digitized businesses during the range of dates where government interventions were initially enforced. The absolute percent increase in national digitization before and after all initial government response is 9\% (from Figure \ref{nationalDigitization}). However, because the United States did not enforce a national mandate, we limit statistical tests in this section to the state-level.

Because of the lack of federal mandates, we examine statewide intervention effects on digitization. For our interrupted time series models, we noticed that all states had an overall significant change in digitization compared to the 2019 control group, with the $\gamma$ median interruption across significant states at 0.65 (corresponding to a relative 93.5\% increase of digitizing merchants); an overview of statewide variations can be seen in Figure \ref{stateVariation}. This digitization occurs \textit{very} quickly: our ITS regression models indicate that all increases in significant states occur within a timeframe of 3 weeks following interventions, then stabilize.

These quantitative results align with our initial survey findings: a majority of respondents attributed digitization to the presence of an intervention, and were forced to digitize in less than a month. Our ITS model quantifies the significance and speed of this interruption identified in our surveys, finding that the enforcement of government mandates---a result of the COVID-19 pandemic---may cause an increase in digitization. However, the variation between overall statewide digitization ($\gamma$: $0.65 \pm .15$ std, also seen in Figure \ref{stateVariation}) indicates that other underlying factors behind government intervention are affecting the adoption of digital payment techniques. We explore these interactions (between state-level factors and digitization) in the next section.  

\begin{table}[]

\begin{tabular}{lrrrrr}
\toprule
                                               & Coefficient & Std. Err & t & P$> |$t$|$  \\
\midrule
Constant                          &       0.1184  &        0.056     &     2.098  &         0.042      \\
Workplace Closing                 &       0.0040  &        0.024     &     0.165  &         0.870      \\
Restrictions on Gatherings**        &       0.0429  &        0.020     &     2.113  &         0.041      \\
Restrictions on Internal Movement &       0.0194  &        0.017     &     1.171  &         0.248      \\
Facial Coverings                  &       0.0114  &        0.041     &     0.275  &         0.785      \\
Republican Vote Share***             &      -0.1087  &        0.034     &    -3.198  &         0.003      \\
Weighed Population Density**        &      -0.0118  &        0.005     &    -2.562  &         0.014      \\
Pre-mandate Digitization***          &       1.4444  &        0.128     &    11.277  &         0.000      \\
COVID Intensity**                   &       0.0314  &        0.015     &     2.072  &         0.045      \\
\bottomrule
\end{tabular}
    \caption{OLS Regression on Predicting Post-Mandate Digitization Rates After Initial Mandate Enforcement. Adjusted $R^{2} \approx .856$. All interlabel correlations are negligible, except for Republican Vote Share and Weighed Population Density ($R^2 = .63$). ** denotes $p < 0.05$, *** denotes $p < 0.01$ }

    \label{ols_results}
\end{table}

\subsection{Relationships Between State Level Factors and Digitization}
\label{stringency_digit_findings}
To understand how underlying policy intervention factors affect digitization, we use several regional factors to predict statewide increases in digitization using a linear regression model. Of the documented factors (in Table \ref{factorRegressionDefs}), restrictions on gatherings, the political orientation of the state, population density, pre-mandate digitization, and COVID intensity are significant at $p < .05$ in our regression model.

For the restrictions on gatherings variable, the OxCRGT dataset distinguishes the degree of the requirement itself, categorizing government response into 4 levels of varying stringency. An example of the least (1) and most (4) stringent requirements are quoted below.

\begin{quote}
    ``\emph{Beginning March 13, 2020, all public and private gatherings in the State of Illinois of 1,000 or more people are cancelled for the duration of the Gubernational Disaster Proclamation. A public or private gathering does not include normal school or work attendance.}'' -- Illinois, Level 1 Restriction ($>=$ 1000 people)
\end{quote}

\begin{quote}
``\emph{Effective March 28, 2020, at 5:00 P.M., all non-work
related gatherings of 10 persons or more, or non-work related gatherings of any size that cannot
maintain a consistent six-foot distance between persons from different households, are
prohibited.}'' -- Alabama, Level 4 Restriction ($>=$ 10 people)
\end{quote}
According to our linear regression model (Table \ref{ols_results}), if all other factors are held constant, recommending the restriction of gatherings results in a $\approx 1$\% increase in digitization (coef / 4 to account for rescaling). Incrementing the stringency of gathering restrictions results in further $\approx 1\%$ increases in digitization. Interestingly, we find that other restrictions (like restrictions on internal movement) are not significantly correlated with digitization; we suspect that, because restrictions on gatherings generally came earlier than restrictions on movement \cite{hale2020variation}, sellers may have been more proactive in digitizing given signs of the pandemic progressing. When/if states later enforced a stay-at-home requirement, some businesses were already prepared to take payments online. 

Another significant independent variable in our regression is Republican Vote Share from the 2020 election. We suspect that this factor is significant due to urbanization (measured by Weighed Population Density) correlating with the political orientation of a state---urbanized regions may be more receptive to adopting digital payment techniques \cite{mokyr1995urbanization}. Furthermore, precautionary COVID responses appear to be a partisan issue: \citet{gollwitzer2020partisan} finds that in the presence of stay-at-home orders, adherence to social distancing measures is strongly related to political affiliation. Our model reflects a similar finding: a 1\% increase (holding all other factors constant) in Republican Vote Share results in $\approx$ -11\% decrease in digitization.

Finally, facial coverings and workplaces closing do not contribute to the predicted change in digitization. At a state-level, facial coverings were generally not mandated at the onset of the pandemic. For workplaces closing, we expected a decrease in digitization since businesses would be closed entirely; we leave further qualitative and quantitative validation to future work.

\begin{figure}[h]
  \centering
          \includegraphics[width=\linewidth]{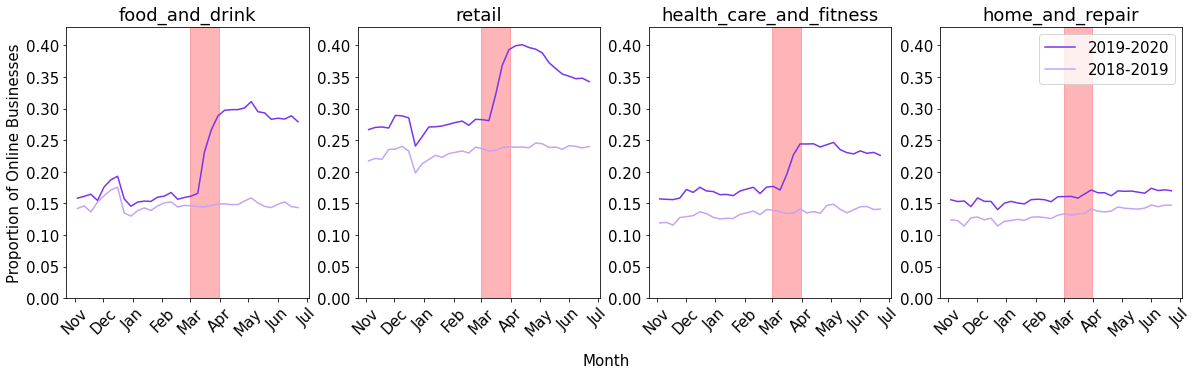}
\caption{The share of businesses that have digitized across business sector in the United States at a weekly level, by \textbf{business sector,} compared between 2018–2019 and 2019–2020. Online sales are based on volume from the same set of sellers across both 2018-2019 and 2019-2020 time spans.}
\label{sectorSegmentationNational}
\end{figure}

\begin{table}[]
\small
\centering
\begin{tabular}{@{}lrrrr@{}}
\toprule
Business Sector & \multicolumn{2}{c}{\# of significant states} & $\mu$ treatment $\gamma$ for all states & Rel. \% $\Delta$ (from $\mu$ treatment $\gamma$) \\ %
\cmidrule{2-3} 
& $> control$ & $< control$ &  &  \\
\midrule
Overall & 50 & 0 & 0.68 & 98.29\% \\
Food \& Drink & 50 & 0 & 1.12 & 207.13\% \\
Retail & 50 & 0 & 0.64 & 88.84\% \\
Health Care \& Fitness & 47 & 0 & 0.61 & 84.90\% \\
Professional Services & 42 & 0 & 0.31 & 35.81\% \\
Home \& Repair & 18 & 0 & 0.16 & 16.88\% \\

\bottomrule
\end{tabular}
\caption{Summary results from ITS model fit across all states. Two-tailed test, $p < .025$, significance compared to control in both directions. Pseudo $R^2 \approx .69$. 
}
\label{stateWideItsResults}
\end{table}

\subsection{Relationships Between Business Category and Digitization} Our survey results from RQ1 suggest a relationship between business sector and challenges associated with digitization. One respondent implied how service sector work requires in-person interaction, limiting the effectiveness and usability of payment digitization. Qualitatively, trends across sectors are visible at the national level: Figure \ref{sectorSegmentationNational} highlights the same payments as Figure \ref{nationalDigitization}, segmented across selected business sectors. Some sectors, like Food \& Drink and Retail, reflect a significant increase in digitization over the span of initial mandates, while other sectors like Professional Services \& Home and Repair highlight a low or reduced share of digitizing businesses. 

Quantifying the significance of digitization at the state level through our ITS models (results in Table \ref{stateWideItsResults}) highlights similar trends. Overall, all states have significantly higher rates of digitization following an initial mandate compared to their 2019 control period. However, the number of significant states, along with increases in digitization, varies substantially with business sector. Food and Drink related businesses, for example, see significant increases in 50 states, while Home and Repair related businesses see significant increases in only 18 states.

Similar to findings discussed in RQ1, our state-level quantitative analyses reinforce hypotheses raised by our survey participants. Service sectors may not be able to digitize effectively due to the mode of the service itself. A noteworthy exception to this trend is the \textit{Healthcare and Fitness} sector; compared to other service oriented sectors (Professional Services and Home \& Repair), \textit{Healthcare and Fitness} sees a substantially higher \% increase. We further discuss reasons behind why these differences exist, along with avenues for future work, in our discussion section (Section \ref{discussionSectionRef}).

\begin{table}[]
\small
\begin{tabular}{p{0.34\linewidth}|p{0.62\linewidth}}
\toprule
Research Question                                               & Summarized Findings \\
\midrule
RQ1: Digitization Causes/Challenges \\ \vspace{-.3em} %
& \vspace{-6.75mm}\begin{itemize}[leftmargin=*]
    \item $46.4\%$ of businesses cited COVID-19 as initial reason to digitize. 
    \item About $80\%$ proactively moved online, in under one month.
    \item $76.4\%$ will continue using digitization adopted during pandemic; however, they are experimenting with the role of digital payments.
    \item Service oriented sectors face more challenges with digitization (consulting, makerspaces, etc.), compared to goods (e.g. retail).
    \item Older customers were apprehensive, citing trust and safety issues; digital gifting norms offer a solution.
    \item Cultural norms associated with cash (in Japan) slowed digitization; however, these norms are shifting because of COVID-19.
\end{itemize} \vspace{-4mm} \\ \midrule 
RQ2: International Digitization \\ \vspace{-.3em} %
& \vspace{-6.75mm}\begin{itemize}[leftmargin=*]
    \item In general, international COVID-19 rates did not correlate with digitization, except for 1st surge---evidence for early digitization.
    \item At minimum, all countries saw a 47\% increase following initial COVID-19 surge, compared to a non-COVID timespan. 
    \item Countries digitized at different rates: JP < AU = GB < US < CA. 
\end{itemize} \vspace{-4mm} \\ \midrule
RQ3: U.S. Digitization Case Study \\ \vspace{-.3em} %
& \vspace{-6.75mm} \begin{itemize}[leftmargin=*]
    \item All states saw significant digitization increases (93.5\% compared to non-COVID timespan) following initial mandates.
    \item Restrictions on gatherings, political orientation, initial COVID-19 intensity positively correlate with state-level digitization.
    \item Different business sectors digitized at significantly different rates: Food \& Drink > Retail > Health Care > Misc. Services 
    \item Health Care \& Fitness (e.g. Telehealth) showed exceptional increases (89\% compared to non-COVID) as a service sector.
\end{itemize} \vspace{-4mm} \\
\bottomrule
\end{tabular}
    \caption{Summary of findings for research questions.}
    \label{summarized_findings}
\end{table}

\section{Discussion and Conclusion}
\label{discussionSectionRef}

In our work, we study how small businesses reacted in response to the COVID-19 pandemic. Specifically, we analyze how these businesses move their services online when adhering to social distancing and government mandates. We contextualize moving online (i.e., digitization) through the use of several strategies, and study how the adoption of these strategies as a whole changed based on both policy and pre-existing social factors. Compared to prior work, we complement a user survey with a large-scale international quantitative analysis of digitization that occurred due to COVID-19, analyzing tens of millions of payments from over 400,000 businesses. From our mixed-method approach, we contribute several findings across our research questions, adding to prior work \cite{lee2021show, benabdallah2021remote, theatresCOVID, mim2021gospels} focusing on understanding the adoption of digitization caused by COVID-19. Our research findings are summarized in Table \ref{summarized_findings}. In this section, we further discuss implications with respect to the surprising scale and speed of business digitization observed in this work. Importantly, these implications may generalize beyond COVID-19---our work highlights a potentially long-lasting shift in how payments are accepted around the world.

\subsection{Small Business Proactiveness and Early Digitization}
Interestingly, we found that \textit{only} early government interventions and early surges in COVID-19 cases correlated with digitization shifts. After these interventions, businesses maintained levels of digitization, regardless of subsequent shifts in COVID-19 rates. From our findings in RQ3, we noticed that restrictions on gatherings---which often preceded stay-at-home mandates---were the only significant restriction ($p < .05$) in our model when predicting digitization. We suspect that early restrictions caused small business to move online, potentially in preparation for more severe government interventions. Proactiveness is also reflected internationally: we noticed that as COVID-19 rates fluctuated around the world, digitization only saw a single inflection point early (March 2020) in the pandemic. Following March 2020 interventions and the initial COVID-19 maxima, businesses stayed digitized even as some regions saw COVID-19 rates decrease substantially. 

\paragraph{Weatherproofing for Future Interventions} Our surveys and interviews expand on our quantitative observation regarding early digitization. Businesses discussed moving online quickly in anticipation of changing interventions. To summarize, about $80\%$ of surveyed businesses transitioned in under a week. Furthermore, interviewed businesses cited digitization as a form of ``weatherproofing;'' because of the uncertainty regarding changing interventions, having any alternative form of accepting payments was critical to remaining open. In the event of future government mandated lockdowns or public health risks, businesses would be well prepared to continue taking payments.      

\paragraph{Experimenting with the Role of Digital Payments After Early Digitization}  Although businesses shifted online early, our quantitative results \textit{do not capture} how businesses continued experimenting with their digital offerings as restrictions varied. For our interviewed and surveyed sellers, however, moving online was an iterative process. Digitization did occur early for these small businesses; however, feedback from customers caused businesses to shift what digitization \textit{techniques} they used. We suspect that given initial usability concerns, businesses experimented (and are still experimenting) with their newly digitized stores. Furthermore, businesses are still in the process of identifying the specific role digital payments play in the context of their in-person offerings.

\subsection{Potential Interactions Between Social Factors and Digitization}

\paragraph{Current Social Norms and Digitization} Another finding of our work highlights potential interactions between pre-existing social norms and the digitization of a region. We find two examples where pre-existing social factors appear to correlate with digitization: political polarization and a cultural reliance on cash. In the United States, political orientation appears to affect digitization: increases in GOP vote share from the 2020 election correlate with decreases in digitization. Prior work also highlights the effect of adherence to mandates (specifically social distancing) in the context of political polarization \cite{gollwitzer2020partisan}. In our work, we show that political orientation transfers to proactive digitization; although digitization itself was never mandated, the adoption of digitization still correlates with the political orientation of a state.

These implications are also reflected internationally. From our brief analysis in Section \ref{beyond_section}, we found that Japan's digitization lagged behind other countries. We suspect that Japan's pre-existing social norms (and correspondingly high cash circulation \cite{jpCashlessPct}) prevented digitization at the same scale as seen in other countries. \citet{japanMeiwaku} qualitatively studies how pre-existing social factors are a potential cause for relatively high use of cash-based transactions (in 2015, only 18\% of payments in Japan were cashless, according to estimates from the World Bank \cite{jpCashlessPct}). 

The specific factors that \citet{japanMeiwaku} studies are \textit{meiwaku} and \textit{tokushita}. To summarize, avoiding \textit{meiwaku} means avoiding commotion: paying by card---especially for a small expense--might require entering a PIN code, or waiting for a transaction to be approved. On the other hand, achieving \textit{tokushita} means achieving some personal gain for a transaction, potentially through a loyalty program. Our work provides quantitative evidence that supports conclusions drawn in \citet{japanMeiwaku} regarding \textit{meiwaku} and \textit{tokushita}; despite surges in COVID-19 cases, digitization in Japan was significantly lower compared to other countries. We suspect that these social norms---intertwined with Japan's dependence on cash---limited overall digitization.

\paragraph{\textbf{Implications}: Shifting Social Norms during COVID-19}
Although pre-existing norms appear to affect digitization, these norms appear to be shifting in response to the pandemic. Despite digitization differences across regions, \textit{every} location studied in our quantitative analyses (RQ2, RQ3) still saw significant increases in business moving online. Furthermore, digitized businesses are not reverting to traditional payment acceptance techniques, even in regions where government interventions were relaxed, or where COVID-19 levels decreased significantly from their peaks.

Another shifting norm---increased online gifting---concerns the usability gap between older and younger generations. In the next section we discuss this usability gap along with its exacerbation in the context of COVID-19; furthermore, we highlight how shifting online payment norms and how specific digitization techniques can address these challenges. 

\subsection{On Digitization's Generational Usability Challenges}
One implication for our work concerns usability and digitization across age groups. Because of the COVID-19 pandemic, some businesses were forced to digitize regardless of usability issues. In these situations, usability problems for customers did not simply disappear; small businesses used a range of techniques to mitigate these issues. One specific usability concern---raised by both our interviewees and survey respondents---relates to trust and safety.

\paragraph{Trust and Safety} For customers, this meant adapting to potentially foreign and ``untrustworthy'' payment methods when businesses decided to go online. Even before the onset of COVID-19, \citet{vinesCheque} highlights how older users were negatively affected by the deprecation of cheques in Britain. From our survey, we find that forced digitization may also result in older customers being apprehensive or left behind. We also find that some businesses highlight customers' distrust of digital payments, exacerbated by the pandemic; this finding also reinforces conclusions from \citet{vinesCheque}, \citet{kaye2014money}, and \citet{smallBusinessSecurity}, who note general security and privacy concerns when moving documentation for personal finances online. Importantly, trust concerns may signal towards a wider mismatch between values of digitization strategies and older users \cite{knowles2018older}. 

From our findings, we worry that these concerns have been realized at a larger scale; in the context of our work, the speed at which businesses moved their operations online alienated users who may have preferred traditional payment mediums. We also suspect that some of these effects are also reflected in Japan's digitization rates. Compared to the remaining countries studied in this work, Japan's small business population is significantly older \cite{economyAndTradeJapan}, and may be averse to digitization. Despite these concerns, interviewed small businesses highlighted shifting norms and experimenting with different digital payment techniques as potential solutions.

\paragraph{\textbf{Implications:} Bridging the Usability Gap} In our survey and interviews, businesses mentioned increasing efforts aimed towards educating older users, following their shifts online. Although directly educating users is a potential solution, businesses also utilized diverse digitization techniques to address trust and safety concerns. One shifting payment norm, highlighted in our qualitative analysis (RQ1), is increased gifting from online stores. Instead of directly purchasing goods or services from an online store, individuals can purchase these items on behalf of others through gifting. Businesses discussed using these digitization techniques to bridge usability gaps across generations. Specifically, younger and more technologically savvy individuals would order/gift items on behalf of older relatives.

Gifting, however, fully removes an individual from a payment transaction loop. To address this, businesses also highlighted the usefulness of online appointments. Older individuals who were hesitant to input information online have the option to schedule a visit, \textit{then} pay in person. These hybrid interactions may serve as a bridge between fully online payments popularized during COVID-19, and traditional in-person interactions that older individuals are more familiar with. Beyond trust and safety implications, emphasizing hybrid digital/physical payment techniques may help customers meet directly with staff, facilitating interpersonal interactions. Enumerating design challenges for older users is an active research area. \citet{coleman2010engaging} and \citet{lindsay2012engaging}, for example, highlight how participatory design sessions can inspire features that engage older individuals. Similarly, these techniques can be extended to online businesses.

\subsection{Facilitating Interpersonal Interactions with Digitized Small Businesses}
A key issue raised by interviewed and surveyed businesses is the lack of interpersonal interactions between customers and business staff when digitizing payments. Though interpersonal interactions are core to the \textit{usability} of payments for older individuals, businesses still highlighted a lack of interaction, intrinsic to building \textit{loyalty} and offering \textit{personalization} to customers. Online stores might have the potential to increase the range of customers visiting a businesses; these weaker ties are undeniably useful, especially for career related support \cite{granovetter1973strength}. However, building stronger ties\footnote{\citet{granovetter1973strength} broadly defines ``strong ties'' as relationships where individuals can directly relay immediate trust and support---through some form of bonding social capital \cite{burt2010neighbor}---to eachother.}---through close relationships between customers and business owners---is a potential challenge. 

\paragraph{\textbf{Implications} Raised by Small Businesses} Through interviews and surveys, we've gleaned potential solutions towards facilitating interactions with staff. Businesses highlighted options like digital podcasts created by staff, or interactive catalogs that allowed users to digitally ``visit'' their physical stores. A key design distinction raised by these businesses---in the context of generational usability---is an emphasis on the interactions with customers and the business itself. For example, gifting might allow older individuals to continue \textit{purchasing} items from digitized businesses; however, these individuals are detached from personalization offered by staff. Although businesses are experimenting with solutions that increase staff-customer interpersonal interactions, these solutions are still in their infancy. Because businesses moved online with unprecedented scale and speed, future research on payment technology systems should continue towards observing and facilitating interactions between staff and customers---especially due to their widespread reduction caused by COVID-19. For example, integrating digital conversational agents \cite{convAgentPersonality} that cater to customer personalities \cite{diyiOnlineRoles, diyiLoans} may provide some personalization lost through physical interpersonal interactions. We suspect that business owners may have had ``strong ties'' with customers before the onset of the COVID-19 pandemic; these ties were threatened upon digitization. This might explain why hybrid interactions are seeing some success; maintaining strong ties relies on a diverse range of communication channels \cite{haythornthwaite2002strong, burke2014growing}. Identifying methods to build and maintain strong ties for small businesses (beyond early-stage hybrid interactions) is an avenue for future work.

\subsection{Digitization and Usability Across Business Sectors}
In contrast to prior work, we also focus on potential usability issues from businesses across different sectors in the United States. To summarize RQ3, while all 50 states in the United States saw significant increases in digitization for Food and Drink sectors, only 18 saw significant increases in online payment methods for Home and Repair sectors. From our surveys, we noticed that service-oriented sectors may have struggled to digitize compared to businesses that sold goods. In one case, we interviewed a business (P4) that pivoted from a service-oriented nature, to one focused on goods. In contrast, one service-oriented sector---Health Care and Fitness---continued to perform better than we expected. From RQ3 (Section \ref{US_analysis}), we found that Health Care and Fitness saw 47 states with significant increases. In these states, the average relative increase in digitization neared $90\%$, significantly higher than professional/home-repair services (Table \ref{stateWideItsResults}). Furthermore, survey respondents noticed that telehealth usage \textit{increased} over the duration of the COVID-19 pandemic. 

\paragraph{\textbf{Implications:} Learning from Health Care \& Fitness' Relative Success} We suspect that the exceptional increases in Health Care, compared to other service-oriented sectors, are due to its amenability to digitization. Telemedicine and eHealth related technologies (subcategories of healthcare and fitness) are generally cost-effective to adopt, having minimal infrastructure requirements; furthermore, providing digital offerings improves usability for customers who cannot afford the cost of health care delivery \cite{de2015cost}. In contrast, Home and Repair services did not need to digitize due to the hands-on nature of their sector; since people are staying at home due to the pandemic, performing home and repair services on your own might simply be more accessible (instead of risking a COVID-19 infection).

Interestingly, \citet{infrastructureTelehealth} reports the presence \textit{and} shift in the dynamics of digitized interpersonal interactions between patient-doctors in the Global South during COVID-19. In our work, interviewed/surveyed service businesses either notice \textit{no} change in payment dynamics, or pivot entirely to a sector where interpersonal interactions are not as critical (e.g. P4 moving from a Makerspace to Retail). Unlike with goods-oriented businesses, maintaining interpersonal interactions between staff and customer are core to accessing the service itself---extending beyond building loyalty or offering personalization. Identifying how service oriented sectors \textit{beyond telehealth} might incorporate interpersonal interactions is an important design implication for digital payment infrastructure. Findings from transitioning physical courses online \cite{remotePandemicLearning} can potentially be applied to sectors that are currently difficult to digitize (Home \& Repair), though future work is necessary. 

\subsection{Implications for Low Income Communities}
Rapid digitization does have its risks. For example, although digitized telehealth services have the potential to provide specialized medical knowledge to rural areas \cite{singh2012towards, miscione2007telemedicine}, these services are sometimes affected by availability issues and technical difficulties \cite{chandwaniTelemed}. Availability issues might be exacerbated in low-income communities that do not have reliable access to the internet. Furthermore, a significant proportion of low-income individuals are unbanked, relying on cash-based transactions \cite{pickens2009banking}. By transitioning to digitized services, businesses are potentially becoming unusable to these individuals. The significant scale of international digitization observed in our work emphasizes these risks. Continuing to accept cash---or designing accessible financial systems that empower unbanked communities to interact with digital payment infrastructure---is critical to ensuring access as more businesses move online. 

\subsection{Future Work}

\paragraph{Understanding Customer Preferences} An avenue for future work involves understanding if \textit{customers} prefer digital or physical sales in regions where businesses have reopened safely. Our work highlights the proportion of businesses that utilize digital techniques to remain open, not customers. Although we suspect that continued digitization of sellers might indicate a consumer preference for digital payments, more work needs to be conducted in this area. As regions approach normalcy, understanding if customers continue preferring digitization in these regions will shed light on changing payment mediums after the pandemic. %

\paragraph{Understanding the Use of Different Digitization Strategies} In this work, we group all digitization techniques together and do not differentiate between, for example, online menus and delivery services. However, we found that the adoption of digitization techniques overall might be dependent on a variety of different factors; therefore, understanding where and why \textit{specific} techniques were successful is important avenue for future work. Because social factors appear to influence digitization, different digitization strategies may be amenable to different pre-existing social norms. %

\paragraph{Permanence of Digital Payments}
Because our preliminary survey respondents indicated that they already had digitization infrastructure in place, we initially suspected that the initial wave of digitization in March and April of 2020 had been driven by businesses that were already on the verge of digitizing anyway. However, we would have expected the rate of digitization to have slowed or decreased from the pre-COVID-19 level. Qualitatively, however, after a small bit of pullback immediately in the months after March and April, the rate of digitizations appears to have largely resumed its gradual rise---well ahead of where it would have been without the pandemic. Although our survey findings indicate that businesses will continue using adopted digital techniques beyond the pandemic, understanding if/how these sentiments change deserves further attention. 

\subsection{Limitations}
\label{discussion_limitations}
\paragraph{Analysing Policy Variations Beyond the US}
Collecting fine-grained policy level data while controlling for federal regulations is a significant, resource intensive challenge. We were able to find a comprehensive dataset for subnational interventions \textit{and corresponding data on small businesses} in only the United States. Other policy datasets were at a national level, or incomplete. Furthermore, as seen in Figure \ref{countryFigureSurvey}, a majority of small businesses in our dataset were located within the United States; separating users into subnational segments from under-covered countries was prohibitive. If these datasets are collected, future work can address these limitations by analyzing how these policies affected digitization, similar to Section \ref{stringency_digit_findings}. 

\paragraph{Finer-Grained Policy Responses in the United States} One limitation of our work is the use of state-level regulations as a proxy for overall government response. However, some city-level responses vary significantly from state-level regulations, affecting digitization in these subregions. Potential future work might take a closer look at a specific set of cities or regions where policy interventions are accessible at a finer resolution. Understanding how different interventions affected digitization in geographically close regions may shed more light on what makes an intervention successful.

\paragraph{Focusing on Businesses That Closed}
Our work focuses on a sample of businesses that remained open throughout the pandemic. Looking at the adoption of digital payment techniques for businesses that have closed permanently may potentially raise a different set of usability and adoption challenges. Identifying these challenges might inspire solutions that help potentially closing businesses remain open. Furthermore, identifying \textit{discrepancies} between strategies adopted by closed and open businesses may help inform struggling businesses. 

\paragraph{Observational Study} 
Finally, although we adopt a quasi-experimental setup and include control variables in our models, our study is still observational; we cannot claim that COVID-19 directly \textit{caused} any of the phenomena documented in our work. However, we carefully considered and included control variables that accounted for temporal and geographic trends, making our work stronger than a purely correlational study.

\section*{Acknowledgements}
We would like to thank the anonymous reviewers for their helpful comments, along with the Commerce ML \& Comms and Policy Teams at Block Inc, and the GT SALT Lab. DY is supported by the Microsoft Research Faculty Fellowship, and by NSF Grant \#2112633 (AI-CARING). 

\bibliographystyle{ACM-Reference-Format}
\bibliography{sample-base}

\appendix
\section{Interview/Survey Questions and Platform Details}
\begin{figure}[H]
  \centering
          \includegraphics[width=.7\linewidth]{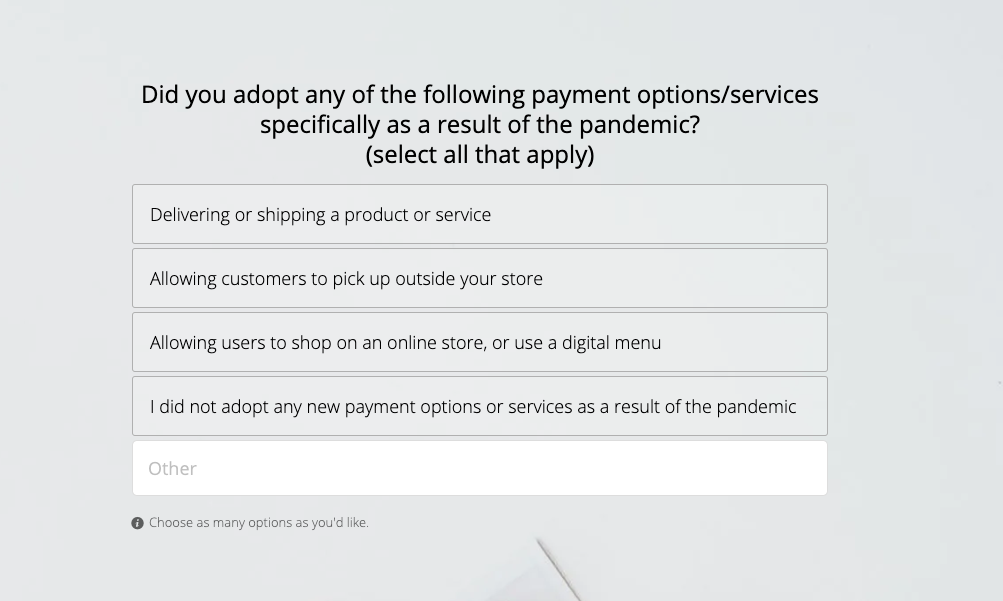}
  \caption{An example question from our survey UI. }
    \label{gtfdbck}
\end{figure}

\paragraph{Survey Platform Details} Our survey was hosted on GetFeedback,\footnote{https://www.getfeedback.com/} a platform used to host online questionnaires. An example of screenshot from GetFeedback's UI can be seen in Figure \ref{gtfdbck}. We delivered email requests directly to registration emails provided by small businesses. Details on how these emails were collected can be found in Section \ref{survey_data_section}.
\label{fullquestionlist}
\paragraph{Survey Questions}
\begin{enumerate}
\item Did the pandemic force you to find ways to take payments online, or in a digital fashion?
\begin{itemize}
    \item Yes, the pandemic forced me to find ways to take payments online or in a digital fashion.
    \item No, the pandemic did not force me to find ways to take payments online or in a digital fashion. 
    \item I already had a digital presence before the pandemic
    \item Other
\end{itemize}

\item Did stay-at-home mandates or government interventions *specifically* force you to find ways to take payments online, or in a contactless fashion?
\begin{itemize}
    \item Yes, stay-at-home mandates forced me to find ways to take payments online or in a contactless fashion.
    \item No, stay-at-home mandates did not force me to find ways to take payments online or in a contactless fashion.
    \item I already had a digital presence before stay-at-home mandates
    \item Other
\end{itemize}

\item Did you adopt any of the following payment options as a result of the pandemic? (select all that apply, or none)

\begin{itemize}
    \item Delivering or shipping a product or service
    \item Allowing customers to pick up outside your store
    \item Allowing users to shop on an online store, or use a digital menu
    \item I did not adopt any new payment options or services as a result of the pandemic
    \item Other
\end{itemize}

\item Which of the following payment options you adopted as a result of the pandemic are you happy with? (select all that apply, or none)

\begin{itemize}
    \item Delivering or shipping a product or service
    \item Allowing customers to pick up outside your store
    \item Allowing users to shop on an online store, or use a digital menu
    \item I did not adopt any new payment options or services as a result of the pandemic
    \item Other
\end{itemize}

\item Which of the following payment options you adopted as a result of the pandemic do you plan to continue to offer when the pandemic is over? (select all that apply, or none)
\begin{itemize}
    \item Delivering or shipping a product or service
    \item Allowing customers to pick up outside your store
    \item Allowing users to shop on an online store, or use a digital menu
    \item I did not adopt any new payment options or services as a result of the pandemic
    \item Other
\end{itemize}
\item If you have any feedback, observations, or challenges relating to business digitization because of the pandemic, please feel free to comment below!
\end{enumerate}

\begin{figure}[H]
  \centering
          \includegraphics[width=.4\linewidth]{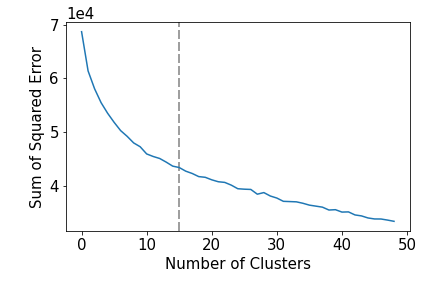}
  \caption{Sum of Squared Error (SSE) for various values of k used by KMeans clustering applied to open-ended survey responses (described in Section \ref{clustering}).}
  \label{clustering_sse}
\end{figure}

\paragraph{Interview Questions}
\begin{enumerate}
    \item \textit{Tell us a bit about you \& your business.}
    \item \textit{Can you recount the events before moving online, and what you did/didn't do when making an online transition?} 
    \item \textit{Were there any parts of digitization in particular you were hesitant about, or anything in particular you found useful?}
    \item \textit{Was there a specific point in time where you shifted? Was your shift online all at once, or was it a more gradual process?}
    \item \textit{Were any customers attracted/alienated by your shift online?} 
    \item \textit{Were there any social factors, specific to your business or location, that affected how you moved online?}
    \item \textit{Did government interventions/support affect how you digitized?}
    \item \textit{Do you plan on continuing to remain digitized after the pandemic? On a related note, do you have any thoughts on the future of digital payments because of the pandemic?}
\end{enumerate}

\end{document}